\documentclass[twocolumn]{aa} 

\usepackage[varg]{txfonts}
\usepackage{graphicx,subfigure,color,amsmath,pifont,multirow,threeparttable}

\usepackage{acronym}
\usepackage{mathrsfs}
\usepackage{gensymb}
\usepackage{hyperref}
\usepackage[mathlines]{lineno}
\usepackage{tabularx}
\usepackage{array}

\begin{document} 

\title{A magnetar-powered jet--wind outflow in gamma-ray burst-associated supernovae: Application to the thermal components of GRB~060218--SN~2006aj}

\author{Yun-Peng Li \inst{\ref{inst1}}
\and Da-Bin Lin \inst{\ref{inst1}}}

\institute{Guangxi Key Laboratory for Relativistic Astrophysics, School of Physical Science and Technology, Guangxi University, Nanning 530004, China\label{inst1}
\\
\email{lindabin@gxu.edu.cn}
}

\date{Received XXX / Accepted XXX}

\abstract{
	A rapidly rotating magnetar formed in the collapse of a massive star may
	release its spin-down power through both a collimated polar jet and a
	quasi-isotropic magnetar wind. With its long-lasting
	soft X-ray blackbody component and the double-peaked UV/optical light curves
	of SN~2006aj, GRB~060218--SN~2006aj provides a useful test case for such a two-component
	energy-injection scenario. We developed a magnetar-powered jet--wind model in
	which the two outflows interact with homologously expanding SN ejecta, driving forward shocks that produce shock-breakout emission and subsequent
	magnetar-powered thermal emission. We first used the jet component to fit the
	$0.3$--$10\,{\rm keV}$ blackbody flux and temperature evolution of GRB~060218 and subsequently
	added the wind component to model the UV/optical light curves of SN~2006aj. We
	find that the soft X-ray blackbody component can be reproduced mainly by
	shock-breakout emission from the polar ejecta driven by the jet. The
	double-peaked UV/optical light curves are instead produced by the interaction
	between the quasi-isotropic magnetar wind and a more massive ejecta component. The first optical peak is dominated by wind-driven shock-breakout emission,
	whereas the broader and later emission is powered mainly by the subsequent
	magnetar-powered thermal component. Our results suggest a two-component
	magnetar-engine picture for GRB~060218--SN~2006aj, in which the X-ray thermal
	emission and the UV/optical double-peaked evolution arise from different
	energy-injection channels of the same central engine.
}


\titlerunning{Magnetar jet--wind model for GRB~060218--SN~2006aj}
\authorrunning{Yun-Peng Li et al.}

\maketitle
\nolinenumbers

\section{Introduction}\label{sec:intro}

The collapse of a massive star may leave behind either a black hole or a neutron-star (NS) remnant (\citealp{1992ApJ...392L...9D,1992Natur.357..472U,1993ApJ...405..273W}).
The extraction of rotational energy or accretion energy from such compact remnants is widely considered to be a possible power source for long-duration gamma-ray bursts (GRBs) and their associated broad-lined type Ic supernovae (SNe Ic-BL; \citealt{1993ApJ...405..273W,1999ApJ...524..262M,2004ApJ...611..380T,2006ARA&A..44..507W,2011MNRAS.413.2031M}). If the remnant is a rapidly rotating, highly magnetized NS -- namely a millisecond magnetar -- its spin-down power can be channelled into both a collimated, Poynting-flux-dominated jet and a quasi-isotropic magnetar wind (\citealp{2004ApJ...611..380T,2009MNRAS.396.2038B,2011MNRAS.413.2031M,2015MNRAS.454.3311M}). The former may launch the relativistic GRB outflow, whereas the latter can deposit energy into the surrounding SN ejecta and modify the thermal emission (\citealp{2009MNRAS.396.2038B,2010ApJ...717..245K,2011MNRAS.413.2031M,2016ApJ...821...36K,2026ApJ..1005..110L}). Sustained or intermittent central-engine activity may also give rise to X-ray plateaus or flares in GRB afterglows (\citealp{2001ApJ...552L..35Z,2005Sci...309.1833B,2006ApJ...642..389N,2006ApJ...642..354Z}). In addition, millisecond magnetars have been proposed as central engines of at least a subset of superluminous supernovae (SLSNe), whose peak luminosities are typically one to two orders of magnitude higher than those of ordinary core-collapse SNe (\citealp{2010ApJ...717..245K,2010ApJ...719L.204W,2012Sci...337..927G,2013ApJ...770..128I,2017ApJ...840...12Y}).

Since the discovery of the GRB~980425 and SN~1998bw association (\citealp{1998Natur.395..672I,1998Natur.395..670G}), a growing sample of GRB-associated supernovae (GRB-SNe) has been identified (\citealp{2006ARA&A..44..507W,2018ApJ...862..130L,2025A&C....5200954F}).
These events are usually classified as SNe Ic-BL and show broad absorption features, high photospheric velocities, and large kinetic energies compared with ordinary stripped-envelope core-collapse SNe (\citealp{2006ARA&A..44..507W,2014MNRAS.443...67M,2016ApJ...832..108M,2017AdAst2017E...5C}).
Although GRB-SNe are generally less luminous than SLSNe, their high ejecta velocities and large kinetic energies may indicate an additional central-engine contribution to the explosion. In this context, millisecond magnetars provide a natural framework for interpreting GRB-SNe: part of the spin-down power can be carried by a relativistic jet that produces the GRB emission, while the remaining energy can be deposited into the SN ejecta through a quasi-isotropic outflow (\citealp{2004ApJ...611..380T,2009MNRAS.396.2038B,2011MNRAS.413.2031M,2015MNRAS.454.3311M,2018ApJ...862..130L}).
This picture may provide a useful framework for connecting the GRB emission with the broad-lined type Ic spectra, high ejecta velocities, and large kinetic energies observed in GRB-SNe.

GRB~060218--SN~2006aj is one of the best-studied nearby GRB-SN events.
GRB~060218 was detected by the \textit{Swift} satellite on 2006 February 18 and is characterized by its long duration, low luminosity, and soft X-ray prompt emission with a thermal blackbody component (\citealp{2006Natur.442.1008C,2006Natur.442.1011P,2006Natur.442.1014S}).
Owing to its nearby redshift of $z\simeq0.0335$ and long-lasting emission (\citealp{2006GCN..4786....1C,2006ApJ...643L..99M}), this event and its associated SN, SN~2006aj, were followed in detail over a broad range of wavelengths. The UV/optical light curves of SN~2006aj show a double-peaked structure (\citealp{2006Natur.442.1008C,2006A&A...454..503S,2006A&A...457..857F}), making this event an important laboratory for studying the connection between low-luminosity GRBs and engine-powered SNe.
Several observational properties of GRB~060218--SN~2006aj point to a relatively low-mass progenitor and have motivated central-engine interpretations involving a newly born NS. The progenitor mass has been estimated to be approximately $20\,M_\odot$, suggesting that the collapse may have left behind a NS remnant rather than a black hole (\citealp{2006Natur.442.1018M}). This interpretation is also consistent with the possible production of a relatively large mass ($\sim0.05\,M_\odot$) of neutron-rich stable $^{58}{\rm Ni}$, as inferred from late-time spectral features (\citealp{2007ApJ...658L...5M}). These features make GRB~060218--SN~2006aj a particularly useful event for testing magnetar-powered models of low-luminosity GRB-SNe.

Previous studies have proposed different origins for the emission components of GRB~060218--SN~2006aj. The early soft X-ray thermal component was initially interpreted as a signature of a mildly relativistic shock breakout occurring in a dense stellar wind surrounding the progenitor (\citealp{2006Natur.442.1008C,2007ApJ...667..351W}). However, such an interpretation struggles to simultaneously account for the long duration and energetics of the prompt thermal emission, as well as the early UV/optical excess (\citealp{2007MNRAS.375..240L,2007MNRAS.375L..36G}). An alternative class of models invokes extended, low-mass material around the compact progenitor. In this case, a jet may deposit its energy in the extended material and drive a mildly relativistic shock breakout (\citealp{2015ApJ...807..172N}), or a low-power, long-lived jet -- together with a low-mass circumstellar envelope -- may account for the multiwavelength emission components (\citealp{2016MNRAS.460.1680I}). In these scenarios, early optical emission is generally associated with cooling emission or interactions involving the extended material, whereas the high-energy emission is linked to the breakout or to emission from the low-power jet. Other studies have suggested that part of the early UV/optical or X-ray emission may contain a nonthermal component associated with a low-luminosity, mildly relativistic jet or with its interaction with the surrounding medium (\citealp{2007ApJ...659.1420T,2019MNRAS.484.5484E}).
More recently, \cite{2025MNRAS.542.1269I} propose a non-equilibrium shock-breakout (SBO) scenario for GRB~060218, in which free-free emission may contribute to the early optical excess.

The long-lasting and low-luminosity prompt emission of GRB~060218, together with the double-peaked light curves of SN~2006aj, has also motivated models in which a newly born NS or magnetar acts as the central engine (\citealp{2006Natur.442.1018M,2007ApJ...659.1420T,2022ApJ...936...54Z}).
In particular, \cite{2022ApJ...936...54Z} show that the double-peaked UV/optical light curves of SN~2006aj can be explained by a magnetar-wind-driven forward SBO from the surrounding ejecta, followed by continued magnetar energy injection. However, the spin-down power of a newly born magnetar is not necessarily deposited into a single outflow component. Rather, it may be divided between a collimated, Poynting-flux-dominated jet and a quasi-isotropic magnetar wind (\citealp{2004ApJ...611..380T,2009MNRAS.396.2038B,2011MNRAS.413.2031M,2015MNRAS.454.3311M,2018ApJ...862..130L,2018MNRAS.475.2659M,2019ApJ...877..153Z}).
Motivated by this picture, we developed a combined jet--wind model for GRB~060218--SN~2006aj, in which the magnetar spin-down power is divided between a polar jet and a quasi-isotropic wind. The early X-ray emission of GRB~060218 contains both a nonthermal power-law component and a thermal blackbody component (\citealp{2006Natur.442.1008C,2006Natur.442.1011P,2006Natur.442.1014S,2007ApJ...654..385K}). In the present work, the polar jet component was used to quantitatively model the flux and temperature evolution of the thermal blackbody component, while the possible physical origin of the accompanying nonthermal emission is qualitatively discussed below. The quasi-isotropic wind component, in turn, accounts for the main double-peaked UV/optical evolution through wind-driven SBO emission followed by magnetar-powered thermal emission. The combined model therefore provides a unified magnetar-engine framework for interpreting the early X-ray blackbody emission of GRB~060218 and the double-peaked UV/optical light curves of SN~2006aj.

The paper is organized as follows. In Sect.~\ref{sec2}, we present the magnetar-driven combined jet--wind model and describe the relevant physical processes. In Sect.~\ref{sec3}, we present the fitting results and the corresponding best-fit parameters. Finally, we summarize our results and discuss their implications in Sect.~\ref{sec4}.

\section{Method} \label{sec2}	
\begin{figure*}[!t]
	\sidecaption
	\includegraphics[
	width=12cm,
	trim={0.5cm 4.5cm 0.1cm 0.3cm},
	clip
	]{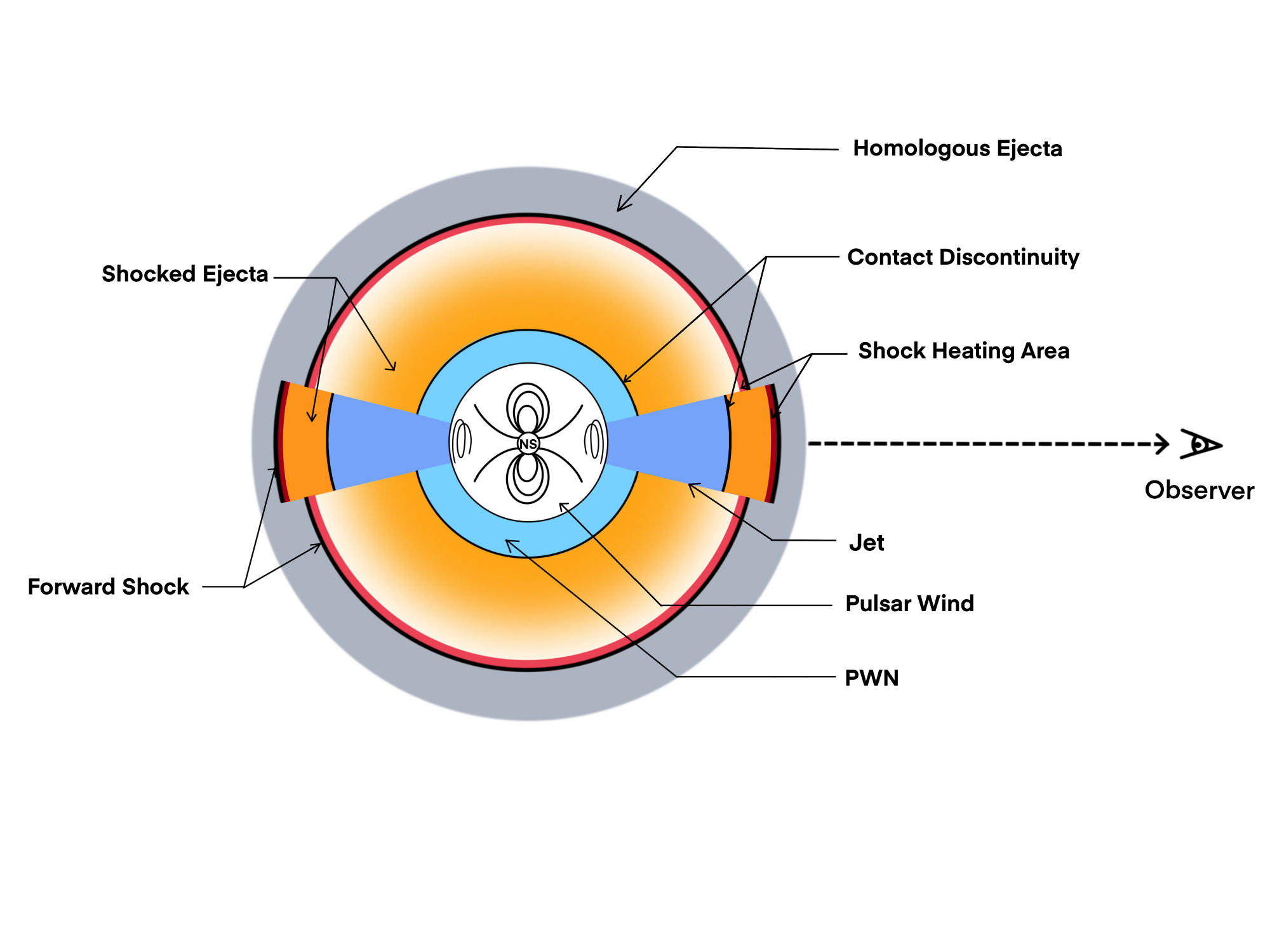}
	\caption{Schematic diagram of the model (not drawn to scale).}
	\label{fig:cartoon}
\end{figure*}

The collapse of a massive star may leave behind a NS remnant surrounded by homologously expanding SN ejecta (\citealp{2003ApJ...591..288H,2005NatPh...1..147W,2009ARA&A..47...63S}). Motivated by previous proto-magnetar jet and magnetar-powered SN models (\citealp{2007MNRAS.382.1029K,2010ApJ...717..245K,2010ApJ...719L.204W,2015MNRAS.454.3311M,2018MNRAS.475.2659M,2019ApJ...877..153Z}), we decomposed the NS spin-down output into two components: a collimated polar jet and a quasi-isotropic relativistic magnetar wind. These outflows interact with the surrounding ejecta and deposit energy into it, driving forward shocks into the ejecta, while the outflow material is terminated by reverse (termination) shocks (\citealp{2022ApJ...936...54Z}).
We separately modelled the radiation powered by these two energy-injection channels and refer to them in this paper as the jet and wind components. Hereafter, the subscripts ${\rm j}$ and ${\rm w}$ denote the quantities associated with the jet and wind components, respectively. A schematic illustration of the model is shown in Fig.~\ref{fig:cartoon}, and the free parameters used in the model fitting are summarized in Table~\ref{tab:free_parameters}.

Following standard semi-analytic treatments of magnetar-powered SNe (\citealp{2010ApJ...717..245K,2010ApJ...719L.204W,2016ApJ...821...36K}), we set $t=0$ at the onset of effective magnetar energy injection through the jet and wind components. The time origin of this model is not necessarily identical to the time of core collapse or the initial SN explosion. We neither explicitly resolved the evolution between the explosion and the onset of the modelled energy-injection phase nor introduced an additional delay-time parameter. Accordingly, $L_{\rm sd,j0}$ and $L_{\rm sd,w0}$ denote the effective spin-down luminosities supplied to the jet and wind components, respectively, at the beginning of the modelled injection phase.

\begin{table}
	\caption{Free parameters of the jet--wind model.}
	\label{tab:free_parameters}
	\centering
	\footnotesize
	\renewcommand{\arraystretch}{1.08}
	
	\begin{tabular}{
			@{}
			>{\raggedright\arraybackslash}p{0.57\columnwidth}
			@{\hspace{0.02\columnwidth}}
			>{\centering\arraybackslash}p{0.14\columnwidth}
			@{\hspace{0.015\columnwidth}}
			>{\centering\arraybackslash}p{0.14\columnwidth}
			@{}
		}
		\hline\hline
		Physical quantity & Jet & Wind \\
		\hline
		Inner density-profile index
		& \multicolumn{2}{c}{$s$} \\
		
		Outer density-profile index
		& \multicolumn{2}{c}{$n$} \\
		
		Spin-down timescale
		& \multicolumn{2}{c}{$T_{\rm sd}$} \\
		
		\hline
		
		Effective ejecta mass
		& $M_{\rm ej,j}$
		& $M_{\rm ej,w}$ \\
		
		Initial ejecta kinetic energy
		& $E_{\rm k,j}$
		& $E_{\rm k,w}$ \\
		
		Initial outer ejecta radius
		& $R_{\rm max,j0}$
		& $R_{\rm max,w0}$ \\
		
		Initial injected spin-down luminosity
		& $L_{\rm sd,j0}$
		& $L_{\rm sd,w0}$ \\
		
		Jet core angle
		& $\theta_{\rm c}$
		& -- \\
		
		X-ray colour-correction factor
		& $f_{\rm col,j}$
		& $f_{\rm col,w}$ \\
		
		\hline
	\end{tabular}
	
	\tablefoot{
		The parameters spanning both component columns are shared between the jet
		and wind components. The subscripts ${\rm j}$ and ${\rm w}$ denote
		the jet- and wind-interacting ejecta, respectively. A dash indicates
		that the corresponding parameter is excluded for that component.
	}
\end{table}

\subsection{Jet component} \label{sub2:1}
\subsubsection{The jet structure and SN ejecta} \label{sub2:1:1}
We considered an axisymmetric structured jet discretized into $N_\theta$ independent angular annuli in the range $0 \leq \theta \leq \theta_{\rm max}$. We fixed $\theta_{\rm max}=0.3\,{\rm rad}$ in our calculations.
The solid angle subtended by the $i$-th annulus is
\begin{equation}
	\Delta \Omega_i =
	2\pi \left(\cos\theta_{i-1/2}-\cos\theta_{i+1/2}\right).
\end{equation}
The angular distribution of the spin-down luminosity carried by the jet is specified by a dimensionless weighting function $g_L(\theta)$.
The normalization factor of the angular luminosity profile is given by
\begin{equation}
	\Omega_L =
	\sum_i g_L(\theta_i)\Delta\Omega_i .
\end{equation}
The spin-down luminosity assigned to the $i$-th angular annulus is written as (\citealp{2001ApJ...552L..35Z,2010ApJ...717..245K})
\begin{equation}\label{Lsdt}
	L_{{\rm sd,j},i}(t)
	=
	L_{{\rm sd,j0},i}
	\left(1+\frac{t}{T_{\rm sd}}\right)^{-2},
\end{equation}
where $T_{\rm sd}$ is the characteristic spin-down timescale.
The initial luminosity in the $i$-th annulus is given by
\begin{equation} \label{Lsd}
	L_{{\rm sd,j0},i}
	=
	L_{\rm sd,j0}
	\frac{g_L(\theta_i)\Delta\Omega_i}{\Omega_L}.
\end{equation}
This normalization ensures that
\begin{equation}
	\sum_i L_{{\rm sd,j0},i}
	=
	L_{\rm sd,j0}.
\end{equation}
In our calculations, we adopted a Gaussian structured jet, for which the angle-dependent luminosity profile is
\begin{equation}
	g_L(\theta)
	=
	\exp\left(-\frac{\theta^2}{2\theta_{\rm c}^2}\right),
\end{equation}
where $\theta_{\rm c}$ is the jet core angle.

For the SN ejecta in the jet region, we distributed the total ejecta mass $M_{\rm ej,j}$ and kinetic energy $E_{\rm k,j}$ among the previously defined angular annuli uniformly per unit solid angle. The mass and kinetic energy assigned to the $i$-th annulus are therefore given by
\begin{equation}
	M_{{\rm ej,j},i} = M_{\rm ej,j}
	\frac{\Delta\Omega_i}{\sum \Delta\Omega_i},
	\quad
	E_{{\rm k,j},i} = E_{\rm k,j}
	\frac{\Delta\Omega_i}{\sum \Delta\Omega_i}.
\end{equation}
This prescription ensures that
\begin{equation}
	\sum_i M_{{\rm ej,j},i}=M_{\rm ej,j},
	\quad
	\sum_i E_{{\rm k,j},i}=E_{\rm k,j}.
\end{equation}
In each angular zone, we assumed a homologously expanding ejecta profile with density (\citealp{2018ApJ...868L..24L,2022ApJ...936...54Z})
\begin{equation}
	\rho_i(x,t)
	=
	\rho_{0,i}\eta(x)
	\left[
	\frac{R_{\rm max,j0}-R_{\rm min,j0}}
	{R_{{\rm max,j},i}(t)-R_{{\rm min,j},i}(t)}
	\right]^3 ,
\end{equation}
where $R_{{\rm min,j},i}(t)=R_{\rm min,j0}+\beta_{{\rm min,j},i}ct$
and $R_{{\rm max,j},i}(t)=R_{\rm max,j0}+\beta_{{\rm max,j},i}ct$.
Here, $\beta_{{\rm min,j},i}$ and $\beta_{{\rm max,j},i}$ are the inner and outer ejecta velocities, respectively, and the subscript zero denotes initial values. The dimensionless radial coordinate is defined as
\begin{equation}
	x=\frac{R-R_{{\rm min,j},i}}{R_{{\rm max,j},i}-R_{{\rm min,j},i}} .
\end{equation}
The density profile is described by a broken power law (\citealp{1982ApJ...258..790C,1999ApJ...510..379M,2010ApJ...717..245K,2013MNRAS.435.1520M}),
\begin{equation} \label{Rho eta}
	\eta(x)=
	\begin{cases}
		(x/x_0)^{-s}, & 0<x<x_0,\\
		(x/x_0)^{-n}, & x_0\leq x\leq 1,
	\end{cases}
\end{equation}
where $s<3$ and $n>5$ are the inner and outer density indices, respectively. In our calculations we adopted $x_0=0.1$.
The density normalization in each angular zone is determined by mass conservation (\citealp{2004A&A...427..453V,2018ApJ...868L..24L}). In the limit $R_{\rm min,j0}\ll R_{\rm max,j0}$, it is given by
\begin{equation}
	\rho_{0,i} = \frac{M_{{\rm ej,j},i}}
	{\Delta\Omega_i I_m R_{\rm max,j0}^3},
\end{equation}
where
\begin{equation}
	I_m
	=
	\int_0^1 \eta(x)x^2\,dx
	=
	\frac{x_0^3}{3-s}
	+
	\frac{x_0^n-x_0^3}{3-n}.
\end{equation}

For material inside the ejecta, $0<x\leq 1$, we prescribed the velocity profile as (\citealp{2016ApJ...819..120L,2022ApJ...936...54Z})
\begin{equation}
	\beta_{{\rm ej},i}(x)
	\simeq
	\beta_{{\rm max,j},i}x,
\end{equation}
the corresponding Lorentz factor is
\begin{equation}
	\Gamma_{{\rm ej},i}
	=
	\left(1-\beta_{{\rm ej},i}^2\right)^{-1/2}.
\end{equation}
The value of $\beta_{{\rm max,j},i}$ is determined by the kinetic energy assigned to the $i$-th angular zone (\citealp{2018ApJ...868L..24L}):
\begin{equation}
	E_{{\rm k,j},i}
	=
	\int
	\left(\Gamma_{{\rm ej},i}-1\right)c^2\,dm ,
\end{equation}
with
\begin{equation}
	dm
	=
	M_{{\rm ej,j},i}
	\frac{\eta(x)x^2\,dx}{I_m}.
\end{equation}
Thus, $\beta_{{\rm max,j},i}$ is obtained by solving
\begin{equation}
	\int_0^1
	\left[
	\left(1-\beta_{{\rm max,j},i}^2 x^2\right)^{-1/2}-1
	\right]
	\eta(x)x^2\,dx
	=
	\frac{I_m E_{{\rm k,j},i}}
	{M_{{\rm ej,j},i}c^2}.
\end{equation}
The adopted values of $R_{\rm min,j0}$ and $\beta_{{\rm min,j},i}$ are sufficiently small and therefore do not significantly affect our results.

\subsubsection{Shock dynamics and SBO} \label{sub2:1:2}
The polar jet outflow interacts with the surrounding SN ejecta and drives a forward shock through them. SBO emission occurs when the optical depth of the unshocked ejecta ahead of the forward shock becomes sufficiently small.
Following the Newtonian treatments of \citet{2016ApJ...819..120L} and \citet{2022ApJ...936...54Z}, we extended the dynamical equations to the mildly relativistic regime. In our model, quantities with a prime symbol ($'$) are measured in the comoving frame of the shocked ejecta. Each angular annulus defined above was evolved independently with the same set of dynamical equations.
For the $i$-th annulus, the total energy of the shocked region is written as
\begin{equation}
	E_i
	=
	(\Gamma_i-1)M_{{\rm sh},i}c^2
	+
	\Gamma_{{\rm eff},i}E'_{{\rm int},i},
\end{equation}
where $\Gamma_i$ is the Lorentz factor of the shocked region, $M_{{\rm sh},i}$ is the swept-up ejecta mass, and $E'_{{\rm int},i}$ is its comoving internal energy.
To account for the Lorentz transformation of the comoving internal energy, we followed \citet{2013MNRAS.433.2107N} and defined $\Gamma_{{\rm eff},i}=(\hat{\gamma}\Gamma_i^2-\hat{\gamma}+1)/\Gamma_i$. We adopted an effective adiabatic index $\hat{\gamma}=4/3$, appropriate for radiation-pressure-dominated shocked ejecta and consistent with previous treatments of engine-heated SN ejecta and SBO emission (\citealp{2016ApJ...819..120L,2017MNRAS.466.2633S,2021ApJ...911..142L,2022ApJ...936...54Z}). Under this assumption, the effective Lorentz factor becomes $\Gamma_{{\rm eff},i}=(4\Gamma_i^2-1)/(3\Gamma_i)$.
The evolution of the total energy is given by
\begin{equation}
	\frac{dE_i}{dt}
	=
	L_{{\rm inj},i}
	-
	L_{{\rm rad},i}
	+
	(\Gamma_{{\rm ej},i}-1)c^2
	\frac{dM_{{\rm sh},i}}{dt},
\end{equation}
where $L_{{\rm inj},i}$ and $L_{{\rm rad},i}$ are the injected and radiated luminosities in the $i$-th annulus, respectively. The unshocked homologously expanding ejecta are assumed to be cold, and their pre-existing internal energy is neglected. The final term therefore accounts only for the bulk kinetic energy carried by the newly swept-up ejecta, whose upstream Lorentz factor is $\Gamma_{{\rm ej},i}$. 
From energy conservation, the Lorentz-factor evolution of the shocked region is then obtained as
\begin{equation} \label{Gamma}
	\frac{d\Gamma_i}{dt}
	=
	\frac{
		L_{{\rm inj},i}
		-
		L_{{\rm rad},i}
		-
		(\Gamma_i-\Gamma_{{\rm ej},i})c^2
		\frac{dM_{{\rm sh},i}}{dt}
		-
		\frac{\Gamma_{{\rm eff},i}}{\Gamma_i}
		\frac{dE'_{{\rm int},i}}{dt'}
	}
	{
		M_{{\rm sh},i}c^2
		+
		E'_{{\rm int},i}
		\frac{d\Gamma_{{\rm eff},i}}{d\Gamma_i}
	}.
\end{equation}
Here, the swept-up mass in the $i$-th angular zone evolves as
\begin{equation}
	\frac{dM_{{\rm sh},i}}{dt}
	=
	\Delta\Omega_i R_{{\rm sh},i}^{2}
	\rho_i(R_{{\rm sh},i},t)
	\left(\beta_{{\rm sh},i}-\beta_{{\rm ej},i}\right)c ,
\end{equation}
where $\beta_{{\rm sh},i}$ is the shock-front velocity and $\beta_{{\rm ej},i}$ is the velocity of the unshocked ejecta at $R_{{\rm sh},i}$. The bulk velocity of the shocked shell is $\beta_i=(1-\Gamma_i^{-2})^{1/2}$. The shock velocity is obtained from the relativistic shock jump conditions (\citealp{2017ApJ...834...32S}),
\begin{equation}\label{shock jump}
	\beta_{{\rm sh},i}
	=
	\frac{
		\hat{\gamma}\Gamma_{{\rm ej},i}\Gamma_i^{2}
		\left(\beta_{{\rm ej},i}-\beta_i\right)\beta_i
		-
		(\hat{\gamma}-1)
		\left(\Gamma_{{\rm ej},i}-\Gamma_i\right)
	}{
		\hat{\gamma}\Gamma_{{\rm ej},i}\Gamma_i^{2}
		\left(\beta_{{\rm ej},i}-\beta_i\right)
		-
		(\hat{\gamma}-1)
		\left(\Gamma_{{\rm ej},i}\beta_{{\rm ej},i}-\Gamma_i\beta_i\right)
	}.
\end{equation}
The shock radius then evolves according to $dR_{{\rm sh},i}/dt=\beta_{{\rm sh},i}c$.

The comoving internal energy in the $i$-th angular annulus is divided into two components: the energy accumulated in the shock-heating layer, $E'_{{\rm sh},i}$, and the remaining internal energy, $\tilde{E}'_i=E'_{{\rm int},i}-E'_{{\rm sh},i}$. Their evolution is described as (\citealp{2016ApJ...819..120L,2022ApJ...936...54Z})
\begin{equation}
	\frac{dE'_{{\rm sh},i}}{dt'}
	=
	H'_{{\rm sh},i}
	-
	P'_{{\rm sh},i}
	\frac{d(\epsilon_{\rm sh}V'_i)}{dt'}
	-
	L'_{{\rm sbo},i},
\end{equation}
\begin{equation}
	\frac{d\tilde{E}'_i}{dt'}
	=
	L'_{{\rm inj},i}
	-
	\tilde{P}'_i
	\frac{dV'_i}{dt'}
	-
	L'_{{\rm sn},i}.
\end{equation}
Here, $E'_{{\rm sh},i}$ powers the SBO component, whereas $\tilde{E}'_i$ represents the thermal reservoir that contributes to the SN-like emission. The shock-heating rate is written as $H'_{{\rm sh},i}=\Gamma_i\left(\Gamma_{{\rm rel},i}-1\right)c^2\frac{dM_{{\rm sh},i}}{dt},$ where $\Gamma_{{\rm rel},i}=\Gamma_i\Gamma_{{\rm ej},i}\left(1-\beta_i\beta_{{\rm ej},i}\right)$ is the relative Lorentz factor between the shocked shell and the upstream ejecta. The corresponding pressures are $P'_{{\rm sh},i}=E'_{{\rm sh},i}/(3\epsilon_{\rm sh}V'_i),$ and $\tilde{P}'_i=\tilde{E}'_i/(3V'_i).$
The comoving volume of the shocked region is approximated as $V'_i=\Delta\Omega_i R_{{\rm sh},i}^{3}/(3\Gamma_i).$
Following \citet{2016ApJ...819..120L} and \citet{2022ApJ...936...54Z}, we write $\epsilon_{\rm sh}=V'_{{\rm sh},i}/V'_i$, where $V'_{{\rm sh},i}$ is the effective comoving volume of the thin layer immediately downstream of the forward shock in which the shock-generated internal energy is concentrated.
Since the detailed post-shock structure is not spatially resolved in our semi-analytic model, we fixed $\epsilon_{\rm sh}=10^{-10}$ for all
angular annuli.\footnote{For sufficiently small values of $\epsilon_{\rm sh}$, the calculated light curves numerically converge: varying $\epsilon_{\rm sh}$ from $10^{-20}$ to $10^{-6}$ results in $\max|\Delta L/L|<0.04\%$, demonstrating that the results are insensitive to its precise value within this range. The adopted value therefore serves as a numerical regularization of the thin-shell limit rather than as a physical estimate of the layer thickness.}

The luminosity injected into the $i$-th angular annulus is the sum of the magnetar spin-down power and the radioactive heating rate, i.e. $L_{{\rm inj},i}=L_{{\rm sd,j},i}+L_{{\rm r},i}.$
The spin-down luminosity follows Eq.~(\ref{Lsdt}), with the adopted angular distribution. The radioactive heating rate is given by (\citealp{1982ApJ...253..785A,1994ApJS...92..527N,2008MNRAS.383.1485V})
\begin{equation}
	L_{{\rm r},i}(t)
	=
	M_{{\rm Ni},i}
	\left[
	\left(\epsilon_{\rm Ni}-\epsilon_{\rm Co}\right)
	e^{-t/\tau_{\rm Ni}}
	+
	\epsilon_{\rm Co}e^{-t/\tau_{\rm Co}}
	\right],
\end{equation}
where $M_{{\rm Ni},i}$ is the mass of $^{56}{\rm Ni}$ in the $i$-th angular annulus, $\epsilon_{\rm Ni}=3.90\times10^{10}\,{\rm erg\,s^{-1}\,g^{-1}}$, $\epsilon_{\rm Co}=6.78\times10^{9}\,{\rm erg\,s^{-1}\,g^{-1}}$, $\tau_{\rm Ni}=8.77\,{\rm days}$, and $\tau_{\rm Co}=111.3\,{\rm days}$. Since the ejecta mass in the jet-axis direction is relatively small, the radioactive-heating contribution in these polar annuli is expected to be subdominant and can be neglected to a good approximation.
In our effective one-zone treatment, $L_{{\rm inj},i}$ is interpreted
as locally deposited power. The deposited energy is assumed to be
rapidly thermalized and isotropized in the comoving frame, and we
adopted $L'_{{\rm inj},i}=L_{{\rm inj},i}$ as an effective source-term
prescription.

Finally, the SBO and SN luminosities are given by (\citealp{2016ApJ...819..120L,2022ApJ...936...54Z})
\begin{equation}\label{Lsbo}
	L'_{{\rm sbo},i}
	=
	\frac{
		R_{{\rm max,j},i}^{2} E'_{{\rm sh},i} c
	}{
		\epsilon_{\rm sh} R_{{\rm sh},i}^{3}
		+
		R_{{\rm max,j},i}^{3}
		-
		R_{{\rm sh},i}^{3}
	}
	\left[
	\frac{
		1-\exp[-(\epsilon_{\rm sh}\tau_{{\rm sh},i}+\tau_{{\rm un},i})]
	}{
		\epsilon_{\rm sh}\tau_{{\rm sh},i}+\tau_{{\rm un},i}
	}
	\right],
\end{equation}
and
\begin{equation}\label{Lsn}
	L'_{{\rm sn},i}
	=
	\frac{\tilde{E}'_i c}{R_{{\rm max,j},i}}
	\left[
	\frac{
		1-\exp[-(\tau_{{\rm sh},i}+\tau_{{\rm un},i})]
	}{
		\tau_{{\rm sh},i}+\tau_{{\rm un},i}
	}
	\right]
\end{equation}
respectively, where $\tau_{{\rm sh},i}=\kappa M_{{\rm sh},i}/(\Delta\Omega_i R_{{\rm sh},i}^{2})$ and $\tau_{{\rm un},i}=\int_{R_{{\rm sh},i}}^{R_{{\rm max,j},i}} \kappa\rho_i\,dr$ are the optical depths of the shocked and unshocked ejecta, and $\kappa=0.1\,{\rm cm^2\,g^{-1}}$ is the opacity (\citealp{2015MNRAS.454.3311M,2022ApJ...936...54Z}).
In the optically thick limit, $\epsilon_{\rm sh}\tau_{{\rm sh},i}+\tau_{{\rm un},i}\gg1$, and for $R_{{\rm sh},i}\ll R_{{\rm max,j},i}$, Eq.~(\ref{Lsbo}) reduces to $L'_{{\rm sbo},i}\simeq E'_{{\rm sh},i}/t'_{{\rm diff},i}$, where $t'_{{\rm diff},i}\simeq \left(\epsilon_{\rm sh}\tau_{{\rm sh},i}+\tau_{{\rm un},i}\right)R_{{\rm max,j},i}/c$ is the effective photon-diffusion timescale.
The total comoving-frame radiative luminosity of the $i$-th annulus is
$L'_{{\rm rad},i}=L'_{{\rm sbo},i}+L'_{{\rm sn},i}.$
Assuming locally isotropic emission in the instantaneous comoving frame of the emitting material, the angle-integrated total radiative power is Lorentz invariant, giving the corresponding source-frame luminosity
$L_{{\rm rad},i}=L'_{{\rm rad},i}$ (\citealp{1979rpa..book.....R,2013MNRAS.430.2703C}).\footnote{Here, $L_{{\rm rad},i}$ denotes the total radiative luminosity in the source frame and should not be confused with the observer-frame flux. The Doppler effect and cosmological redshift are included separately when calculating the observed flux.}

To calculate the multiband light curves, we first assigned an effective
photospheric temperature to the thermal emission, defined as
\begin{equation}
	T'_{{\rm eff},i}
	=
	\left[
	\frac{
		L'_{{\rm rad},i}
	}{
		\Delta\Omega_i\sigma_{\rm SB}R_{{\rm ph},i}^{2}
	}
	\right]^{1/4}.
\end{equation}
The photospheric radius, $R_{{\rm ph},i}$, was obtained from the radius at which the outward optical depth reaches
\begin{equation}
	\tau_{{\rm ph},i}
	=
	\int_{R_{{\rm ph},i}}^{R_{{\rm max,j},i}}
	\kappa \rho_i\,dr
	=
	\frac{2}{3}.
\end{equation} 
If $R_{{\rm ph},i} < R_{{\rm sh},i}$, we simply set $R_{{\rm ph},i} = R_{{\rm sh},i}$.
To smoothly suppress thermal radiation in optically thin regions, we introduced a thermalization factor:
\begin{equation}
	f_{{\rm th},i}=1-\exp[-(\tau_{{\rm sh},i}+\tau_{{\rm un},i})].
\end{equation}

The observed radiation was obtained by summing the relativistically boosted blackbody emission from all angular zones. For an on-axis observer, the arrival time from the $i$-th annulus is
\begin{equation}
	t_{{\rm obs},i}
	=
	(1+z)
	\left(
	t-\frac{R_{{\rm ph},i}\mu_i}{c}
	\right),
\end{equation}
where $\mu_i=\cos\theta_i$.
The photon energy in the comoving frame is
\begin{equation}
	h\nu'_i
	=
	\frac{1+z}{\delta_i}h\nu_{\rm obs},
\end{equation}
with the Doppler factor $\delta_i=1/[\Gamma_{i}(1-\beta_{i}\mu_i)]$.

The temperature defined above should be interpreted as an effective
photospheric temperature. Since our grey-opacity treatment does not solve the frequency-dependent thermalization process, the colour temperature inferred from blackbody fits to the observed spectra may differ from this effective temperature, particularly at X-ray energies (\citealp{1992ApJ...393..742E,2008ApJ...683L.135C,2010ApJ...725..904N,2011ApJ...728...63R,2017hsn..book..967W}). We therefore introduced a phenomenological colour-correction, factor $f_{\rm col,j}$. The observer-frame colour temperature associated with the emission from the $i$-th angular annulus is thus given by
\begin{equation}
	T_{{\rm obs},i}=\frac{\delta_i}{1+z}f_{\rm col,j} T'_{{\rm eff},i}.
\end{equation}
This factor accounts for the difference between the colour temperature and the effective temperature caused by scattering-dominated radiative transfer and by the thermalization depth lying below the scattering photosphere (\citealp{1992ApJ...393..742E,2008ApJ...683L.135C,2011ApJ...728...63R}). In sufficiently fast shocks, incomplete thermalization and Comptonization may drive the radiation away from a Planck spectrum and modify both the colour temperature and the low-energy spectral slope
(\citealp{2010ApJ...725..904N,2010ApJ...716..781K,2013ApJ...774...79S,2022ApJ...928..122M,2025MNRAS.542.1269I}).
The thermalization conditions, and hence the local colour correction, may also vary among angular annuli. Our grey treatment does not resolve these processes, and the available X-ray blackbody flux and temperature do not independently constrain such an angular dependence. We therefore adopted a single $f_{\rm col,j}$ in the present fit as a phenomenological description of the annulus-summed thermal spectrum rather than as implying identical local thermalization conditions throughout the jet.

The observed spectral energy flux was thus calculated using a colour-corrected blackbody spectrum as (\citealp{1995ApJ...445..780S,2013ApJ...776L..40Y})
\begin{equation}
	\begin{aligned}
		\nu_{\rm obs}F_{\nu_{\rm obs}}(\nu_{\rm obs},t_{\rm obs})
		&=
		\sum_i
		\frac{\Delta\Omega_i}{4\pi D_L^2}
		\frac{2\pi}{h^3c^2}
		f_{\rm col,j}^{-4}
		f_{{\rm th},i}
		\delta_i^4 R_{{\rm ph},i}^{2}  \\
		&\quad \times
		\frac{
			(h\nu'_i)^4
		}{
			\exp\left[h\nu'_i/(k_{\rm B}f_{\rm col,j}T'_{{\rm eff},i})\right]-1
		},
	\end{aligned}
\end{equation}
where $\nu'_i=(1+z)\nu_{\rm obs}/\delta_i$ is the comoving-frame frequency.
The factor $f_{\rm col,j}^{-4}$ ensures approximate conservation of the
bolometric luminosity when the spectrum is characterized by the colour temperature.

\subsection{Wind component} \label{sub2:2}

The quasi-isotropic magnetar-wind component is treated as an effective spherical interaction between the wind and the surrounding SN ejecta. We therefore used the same dynamical and radiative prescriptions as those introduced for the jet component but replaced the angular solid angle $\Delta\Omega_i$ by $4\pi$ and omitted the angular-zone index.

The spin-down luminosity injected into the wind component is written as (\citealp{2001ApJ...552L..35Z,2010ApJ...717..245K})
\begin{equation}
	L_{\rm sd,w}(t)
	=
	L_{\rm sd,w0}
	\left(1+\frac{t}{T_{\rm sd}}\right)^{-2},
\end{equation}
where $L_{\rm sd,w0}$ is the initial spin-down luminosity assigned to the wind component, and the same characteristic spin-down timescale $T_{\rm sd}$ as in the jet component was adopted. The total injected luminosity is
\begin{equation}
	L_{\rm inj,w}
	=
	L_{\rm sd,w}
	+
	L_{\rm r},
\end{equation}
where the radioactive heating rate is (\citealp{1982ApJ...253..785A,1994ApJS...92..527N,2008MNRAS.383.1485V})
\begin{equation}
	L_{\rm r}(t)
	=
	M_{\rm Ni}
	\left[
	(\epsilon_{\rm Ni}-\epsilon_{\rm Co})e^{-t/\tau_{\rm Ni}}
	+
	\epsilon_{\rm Co}e^{-t/\tau_{\rm Co}}
	\right],
\end{equation}
where the mass of $^{56}{\rm Ni}$, $M_{\rm Ni}$, was set to $0.1\,M_\odot$.

The ejecta interacting with the wind are assumed to be homologously expanding and approximately spherical. We adopted the same functional form of the broken power-law density profile as in the jet component but allowed the ejecta mass $M_{\rm ej,w}$, kinetic energy $E_{\rm k,w}$, and initial outer radius $R_{\rm max,w0}$ to differ from those in the polar jet region. Their density profile is (\citealp{2016ApJ...819..120L,2022ApJ...936...54Z})
\begin{equation}
	\rho_{\rm w}(x,t)
	=
	\rho_{{\rm w},0}\eta(x)
	\left[
	\frac{R_{\rm max,w0}-R_{\rm min,w0}}
	{R_{\rm max,w}(t)-R_{\rm min,w}(t)}
	\right]^3 ,
\end{equation}
where the broken power-law function $\eta(x)$ is the same as that in Eq.~(\ref{Rho eta}). The density normalization is given by
\begin{equation}
	\rho_{{\rm w},0}
	=
	\frac{M_{\rm ej,w}}
	{4\pi I_m R_{\rm max,w0}^{3}}.
\end{equation}
The maximum ejecta velocity $\beta_{\rm max,w}$ is determined from the kinetic energy of the wind-interacting ejecta,
\begin{equation}
	\int_0^1
	\left[
	(1-\beta_{\rm max,w}^{2}x^{2})^{-1/2}-1
	\right]
	\eta(x)x^2\,dx
	=
	\frac{I_mE_{\rm k,w}}{M_{\rm ej,w}c^2}.
\end{equation}

The Lorentz-factor evolution of the shocked region was calculated using Eq.~(\ref{Gamma}), with all angular-zone subscripts omitted. The swept-up mass evolves as (\citealp{2016ApJ...819..120L,2022ApJ...936...54Z})
\begin{equation}
	\frac{dM_{\rm sh}}{dt}
	=
	4\pi R_{\rm sh}^{2}
	\rho_{\rm w}(R_{\rm sh},t)
	(\beta_{\rm sh}-\beta_{\rm ej,w})c .
\end{equation}
The shock radius evolves according to $dR_{\rm sh}/dt=\beta_{\rm sh}c$, where $\beta_{\rm sh}$ was obtained from Eq.~(\ref{shock jump}).
The comoving internal energy was divided into the shock-heated component $E'_{\rm sh}$ and the remaining thermal component $\tilde{E}'=E'_{\rm int}-E'_{\rm sh}$. Their evolution follows (\citealp{2016ApJ...819..120L,2022ApJ...936...54Z})
\begin{equation}
	\frac{dE'_{\rm sh}}{dt'}
	=
	H'_{\rm sh}
	-
	P'_{\rm sh}
	\frac{d(\epsilon_{\rm sh}V')}{dt'}
	-
	L'_{\rm sbo,w},
\end{equation}
and
\begin{equation}
	\frac{d\tilde{E}'}{dt'}
	=
	L'_{\rm inj,w}
	-
	\tilde{P}'
	\frac{dV'}{dt'}
	-
	L'_{\rm sn,w}.
\end{equation}
Here, $V'=4\pi R_{\rm sh}^{3}/(3\Gamma)$ is the comoving volume of the shocked ejecta, $P'_{\rm sh}=E'_{\rm sh}/(3\epsilon_{\rm sh}V')$, and $\tilde{P}'=\tilde{E}'/(3V')$.

The SBO and SN luminosities were calculated using the same diffusion prescription as in Eqs.~(\ref{Lsbo}) and~(\ref{Lsn}),
\begin{equation}
	L'_{\rm sbo,w}
	=
	\frac{
		R_{\rm max,w}^{2}E'_{\rm sh}c
	}{
		\epsilon_{\rm sh}R_{\rm sh}^{3}
		+
		R_{\rm max,w}^{3}
		-
		R_{\rm sh}^{3}
	}
	\left[
	\frac{
		1-\exp[-(\epsilon_{\rm sh}\tau_{\rm sh}+\tau_{\rm un})]
	}{
		\epsilon_{\rm sh}\tau_{\rm sh}+\tau_{\rm un}
	}
	\right],
\end{equation}
and
\begin{equation}
	L'_{\rm sn,w}
	=
	\frac{\tilde{E}'c}{R_{\rm max,w}}
	\left[
	\frac{
		1-\exp[-(\tau_{\rm sh}+\tau_{\rm un})]
	}{
		\tau_{\rm sh}+\tau_{\rm un}
	}
	\right].
\end{equation}
The optical depths of the shocked and unshocked ejecta are $\tau_{\rm sh}=\kappa M_{\rm sh}/4\pi R_{\rm sh}^{2}$ and $\tau_{\rm un}=\int_{R_{\rm sh}}^{R_{\rm max,w}} \kappa\rho_{\rm w}\,dr$, respectively.
The total comoving radiative luminosity is $L'_{\rm rad,w}=L'_{\rm sbo,w}+L'_{\rm sn,w}.$
The comoving effective temperature is
\begin{equation}
	T'_{\rm eff,w}
	=
	\left[
	\frac{
		L'_{\rm rad,w}
	}{
		4\pi\sigma_{\rm SB}R_{\rm ph,w}^{2}
	}
	\right]^{1/4},
\end{equation}
where the photospheric radius is determined by
\begin{equation}
	\int_{R_{\rm ph,w}}^{R_{\rm max,w}}
	\kappa\rho_{\rm w}\,dr=\frac{2}{3}.
\end{equation}
If $R_{\rm ph,w}<R_{\rm sh}$, we set $R_{\rm ph,w}=R_{\rm sh}$. We also included a thermalization factor,
\begin{equation}
	f_{\rm th,w}
	=
	1-\exp[-(\tau_{\rm sh}+\tau_{\rm un})].
\end{equation}

For the wind component, we calculated the observed spectrum using a blackbody approximation with the same Doppler correction as in the jet component. In the one-zone spherical approximation, the arrival time is
\begin{equation}
	t_{{\rm obs},{\rm w}}
	=
	(1+z)
	\left(
	t-\frac{R_{\rm ph,w}}{c}
	\right).
\end{equation}
The comoving photon energy is
\begin{equation}
	h\nu'_{\rm w}
	=
	\frac{1+z}{\delta_{\rm w}}h\nu_{\rm obs},
\end{equation}
where $\delta_{\rm w}=1/[\Gamma(1-\beta)]$ is the Doppler factor for the line-of-sight photospheric material and $\beta=(1-\Gamma^{-2})^{1/2}$.
Analogously to the jet component, we introduced a colour-correction factor $f_{\rm col,w}$ to account for the difference between the colour temperature and the effective temperature of the wind component, such that
\begin{equation}
	T_{{\rm obs},{\rm w}}
	=
	\frac{\delta_{\rm w}}{1+z}
	f_{\rm col,w} T'_{\rm eff,w}.
\end{equation}
The observed spectral energy flux is hence
\begin{equation}
	\begin{aligned}
		\nu_{\rm obs}F_{\nu_{\rm obs},{\rm w}}(\nu_{\rm obs},t_{{\rm obs},{\rm w}})
		&=
		\frac{1}{4\pi D_L^2}
		\frac{8\pi^2}{h^3c^2}
		f_{\rm col,w}^{-4} f_{\rm th,w}
		\delta_{\rm w}^{4}
		R_{\rm ph,w}^{2}  \\
		&\quad \times
		\frac{
			(h\nu'_{\rm w})^{4}
		}{
			\exp\left[h\nu'_{\rm w}/(k_{\rm B}f_{\rm col,w}T'_{\rm eff,w})\right]-1
		}.
	\end{aligned}
\end{equation}

\section{Results} \label{sec3}
\subsection{Best-fit model and dynamical and radiative evolution}
\subsubsection{Model fitting and parameter constraints}
We applied the combined jet--wind model to the thermal X-ray and UV/optical emission of GRB~060218. The observed X-ray emission is dominated by a broad nonthermal continuum, whereas the blackbody component considered here is subdominant. During the early XRT observations, the blackbody contributes approximately $15$--$20\%$ of the observed $0.3$--$10\,{\rm keV}$ flux (\citealp{2006Natur.442.1008C,2007MNRAS.375L..36G}). 
Its mean contribution during the prompt phase is only approximately $0.13\%$ when evaluated over the broader $0.5$--$150\,{\rm keV}$ band (\citealp{2007ApJ...654..385K}). Our model is therefore not intended to reproduce the total high-energy emission. The X-ray fit was restricted to the blackbody flux and temperature inferred from the power-law-plus-blackbody decomposition of the XRT spectra, using the observational constraints reported by \citet{2006Natur.442.1008C} and \citet{2015ApJ...807..172N}.

An additional uncertainty arises from the host-galaxy extinction adopted for the UV/optical data. \citet{2006Natur.442.1008C} and \citet{2015ApJ...807..172N} adopted relatively high host reddening, for which the extinction-corrected UV/optical spectrum is approximately consistent with the Rayleigh--Jeans tail of a hot blackbody, $F_\nu\propto\nu^2$. By contrast, \citet{2006A&A...454..503S} favoured a substantially lower host extinction. More recently, \citet{2025MNRAS.542.1269I} find that adopting lower extinction produces a considerably flatter intrinsic spectrum, with $F_\nu\propto\nu$ around $3\times10^4\,{\rm s}$ and an even flatter spectrum at earlier times. The wind-component fit presented here therefore depends on the adopted high-extinction correction. For the lower extinction, the corrected UV/optical spectrum deviates substantially from the Rayleigh--Jeans behaviour assumed in the present fit. Reproducing the corresponding spectral shape would likely require an additional non-blackbody emission component or a more detailed radiative-transfer treatment.

\begin{figure}[!t]
	\centering
	\includegraphics[width=\columnwidth]{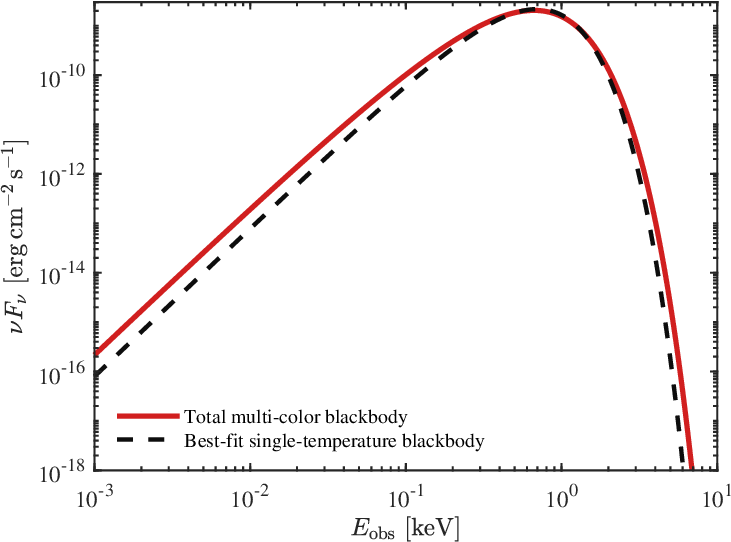}
	
	\caption{
		Comparison between the total multi-colour blackbody spectrum of the jet
		component (solid red line) and a single-temperature blackbody spectrum
		(dashed black line) at $t_{\rm obs}=10^{3}\,{\rm s}$.
		The multi-colour spectrum was obtained by summing the emission from all
		angular annuli, whereas the single-temperature blackbody has
		$kT_{\rm BB}=0.17\,{\rm keV}$.
		The two spectra agree closely around the spectral peak, while the
		multi-colour spectrum is slightly broader towards both lower and higher
		energies.
	}
	
	\label{multicolor spectrum}
\end{figure}

The model parameters were constrained using Markov chain Monte Carlo (MCMC) sampling with the \texttt{emcee} package (\citealp{2013PASP..125..306F}). We first fitted the $0.3$--$10\,{\rm keV}$ flux of the X-ray blackbody component together with its temperature evolution using only the jet component. In the jet calculation, the outflow was divided into axisymmetric angular annuli, each characterized by its own photospheric radius, Doppler factor, and effective temperature. To compare with the observed blackbody temperature, we summed the observer-frame spectra of all annuli and fitted the resulting multi-temperature spectrum with a single-temperature blackbody over the $0.3$--$10\,{\rm keV}$ band. The resulting temperature should therefore be interpreted as an equivalent observer-frame blackbody temperature rather than the temperature of any individual annulus. The corresponding parameter constraints are summarized in Table~\ref{tab:grb060218_jet_fit}. 
The corresponding summed multi-colour spectrum is shown in Fig.~\ref{multicolor spectrum}. It remains close to the best-fitting single-temperature blackbody around the spectral peak. At the epoch shown, the corresponding equivalent blackbody temperature is $kT_{\rm BB}=0.17\,{\rm keV}$. The distribution of temperatures and Doppler factors among the angular annuli introduces only modest spectral broadening, with the largest deviations from the single-temperature blackbody appearing at low and high energies.

We subsequently fixed the best-fit jet component obtained from the X-ray blackbody fit and added a wind component to reproduce the UV/optical light curves of GRB~060218. Motivated by possible ejecta anisotropies (\citealp{2006ApJ...645.1331M,2006A&A...459L..33G,2007A&A...475L...1M,2014ApJ...785L..29M,2023MNRAS.522.6070P}), we allowed the ejecta interacting with the wind to differ from the polar ejecta interacting with the jet. The wind component therefore has its own ejecta mass $M_{\rm ej,w}$, kinetic energy $E_{\rm k,w}$, initial outer radius $R_{\rm max,w0}$, and colour-correction factor $f_{\rm col,w}$. The radial density indices, $s$ and $n$, and the spin-down timescale, $T_{\rm sd}$, were fixed to the best-fit values obtained from the jet fit. The corresponding parameter constraints are listed in Table~\ref{tab:grb060218_wind_fit}.

\begin{table}[!t]
	\centering
	\caption{Best-fit and posterior parameter estimates for the jet-component model of GRB~060218.}
	\label{tab:grb060218_jet_fit}
	\footnotesize
	\renewcommand{\arraystretch}{1.25}
	\setlength{\tabcolsep}{4pt}
	\begin{tabular}{@{}lccc@{}}
		\hline\hline
		Parameter & Value & Best fit & Prior \\
		\hline
		
		$s$ 
		& $1.663^{+0.44}_{-0.46}$ 
		& $2.086$ 
		& $(0,3)$ \\
		
		$n$ 
		& $5.257^{+0.42}_{-0.19}$ 
		& $5.149$ 
		& $(5,10)$ \\
		
		$\log_{10}(E_{\rm k,j}/{\rm erg})$ 
		& $50.480^{+0.13}_{-0.19}$ 
		& $50.519$ 
		& $(49,52)$ \\
		
		$\log_{10}(T_{\rm sd}/{\rm day})$ 
		& $-2.909^{+0.11}_{-0.07}$ 
		& $-2.978$ 
		& $(-3,0)$ \\
		
		$\log_{10}(L_{\rm sd,j0}/{\rm erg\,s^{-1}})$ 
		& $49.330^{+0.19}_{-0.18}$ 
		& $49.285$ 
		& $(47,50)$ \\
		
		$\log_{10}(M_{\rm ej,j}/M_\odot)$ 
		& $-1.466^{+0.20}_{-0.18}$ 
		& $-1.522$ 
		& $(-3,0)$ \\
		
		$\log_{10}(R_{\rm max,j0}/{\rm cm})$ 
		& $13.210^{+0.07}_{-0.06}$ 
		& $13.231$ 
		& $(12,14)$ \\
		
		$f_{\rm col,j}$ 
		& $1.81^{+0.16}_{-0.10}$ 
		& $1.829$ 
		& $(1,3)$ \\
		
		$\theta_{\rm c}$ 
		& $0.061^{+0.017}_{-0.007}$ 
		& $0.058$ 
		& $(0.05,0.2)$ \\

		\hline
	\end{tabular}
	\tablefoot{
		The estimates were obtained from an MCMC fit to the X-ray blackbody light curve and the corresponding blackbody-temperature evolution.
	}
\end{table}

\begin{table}[t]
	\centering
	\caption{Parameters estimated from fitting the multiwavelength emission of GRB~060218 with the combined jet--wind model.}
	\label{tab:grb060218_wind_fit}
	\footnotesize
	\renewcommand{\arraystretch}{1.25}
	\setlength{\tabcolsep}{4pt}
	\begin{tabular}{lccc}
		\hline\hline
		Parameter & Value & Best fit & Prior \\
		\hline
		
		$\log_{10}(E_{\rm k,w}/{\rm erg})$ 
		& $50.420^{+0.02}_{-0.02}$ 
		& $50.420$ 
		& $(49,52)$ \\
		
		$\log_{10}(L_{\rm sd,w0}/{\rm erg\,s^{-1}})$ 
		& $50.500^{+0.00}_{-0.00}$ 
		& $50.500$ 
		& $(47,50.5)$ \\
		
		$\log_{10}(M_{\rm ej,w}/M_\odot)$ 
		& $0.685^{+0.01}_{-0.01}$ 
		& $0.683$ 
		& $(0,1)$ \\
		
		$\log_{10}(R_{\rm max,w0}/{\rm cm})$ 
		& $13.910^{+0.01}_{-0.01}$ 
		& $13.912$ 
		& $(12,14)$ \\
		
		$f_{\rm col,w}$ 
		& $1.000^{+0.00}_{-0.00}$ 
		& $1.000$ 
		& $(1,3)$ \\

		\hline
		\multicolumn{4}{l}{\text{Fixed parameters}} \\
		\hline
		
		$s$ 
		& $2.086$ 
		& fixed 
		& -- \\
		
		$n$ 
		& $5.149$ 
		& fixed 
		& -- \\
		
		$\log_{10}(T_{\rm sd}/{\rm day})$ 
		& $-2.978$ 
		& fixed 
		& -- \\
		
		\hline
	\end{tabular}
	\tablefoot{
		The fit used MCMC sampling. The fixed parameters were taken from the best-fit jet-component model.
	}
\end{table}

\subsubsection{Dynamical and radiative evolution}
\begin{figure*}[!t]
	\centering
	\includegraphics[width=0.97\textwidth]{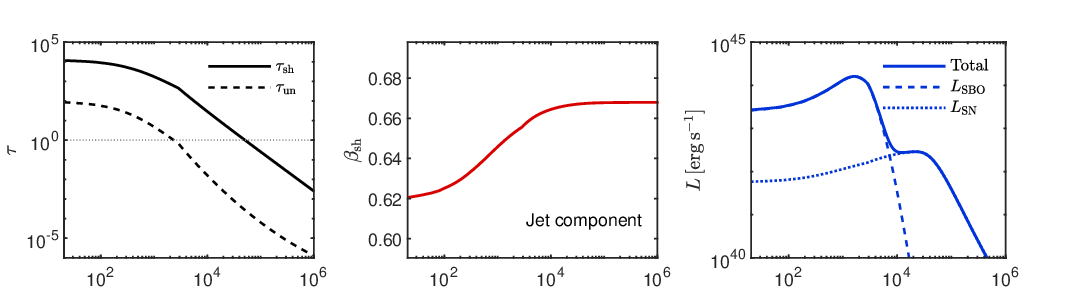}
	\includegraphics[width=0.97\textwidth]{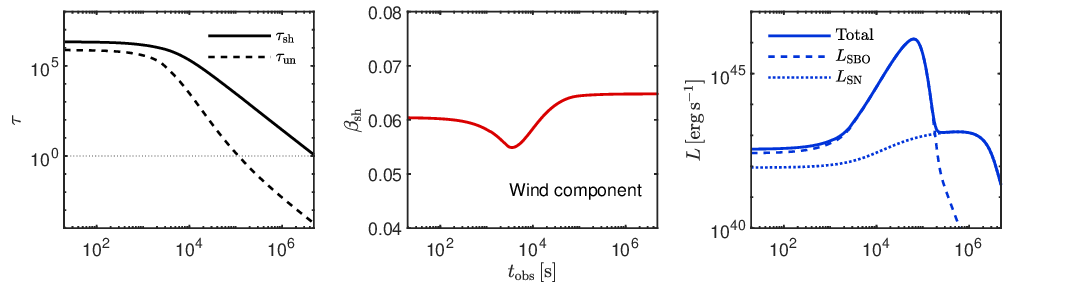}
	\caption{
		Evolution of the optical depths (left column), forward-shock velocity (middle column), and bolometric luminosity (right column) of the jet (top row) and wind (bottom row) components, calculated using the corresponding best-fit parameters. In the left panels, the solid and dashed curves show the optical depths of the shocked and unshocked ejecta,
		respectively. In the right panels, the dashed and dotted curves represent the SBO and SN luminosities, respectively, while the solid curves show their sum. For the jet component, the results are shown for the innermost angular annulus closest to the jet axis, whereas the wind component is treated as an effectively isotropic outflow.
	}
	
	\label{dynamic evolution}
\end{figure*}

Figure~\ref{dynamic evolution} shows the evolution of the optical depths,
forward-shock velocity, and bolometric luminosity for the best-fit jet and
wind components listed in Tables~\ref{tab:grb060218_jet_fit} and
\ref{tab:grb060218_wind_fit}, respectively. For the jet component, we show the innermost angular annulus as a representative example, whereas the wind component is treated as effectively isotropic. The upper and lower rows correspond to the jet and wind components, respectively, while the left, middle, and right columns show the optical depths, forward-shock velocity, and bolometric luminosity.
In both components, the optical depth of the shocked ejecta, $\tau_{\rm sh}$, remains larger than that of the unshocked ejecta, $\tau_{\rm un}$, and decreases more slowly as the forward shock approaches the outer ejecta boundary and the ejecta expand. For the jet component, $\tau_{\rm un}$ becomes of order unity at $t_{\rm obs}\sim10^{3}\,{\rm s}$, whereas this occurs at $t_{\rm obs}\sim10^{5}\,{\rm s}$ for the wind component. Although $\tau_{\rm sh}>\tau_{\rm un}$ over much of the evolution, the escape of the shock-generated radiation depends on the effective optical depth $\epsilon_{\rm sh}\tau_{\rm sh}+\tau_{\rm un}$. Because $\epsilon_{\rm sh}\ll1$, the breakout evolution is controlled primarily by the rapid decrease in $\tau_{\rm un}$. By contrast, the SN radiation diffuses through the total optical depth $\tau_{\rm sh}+\tau_{\rm un}$, which remains large for substantially longer. This leads to the more gradual rise and later release of the SN component.

For the jet component, the shock velocity increases gradually from $\beta_{\rm sh}\simeq0.62$ to $\beta_{\rm sh}\simeq0.67$, so that the forward shock propagates through the polar ejecta at mildly relativistic velocities. The bolometric luminosity is dominated at early times by the SBO component, which reaches a broad maximum of order $10^{44}$--$10^{45}\,{\rm erg\,s^{-1}}$ at $t_{\rm obs}\sim10^{3}$--$10^{4}\,{\rm s}$ and then declines rapidly. The magnetar-powered SN component is initially subdominant but evolves on a longer timescale and becomes relatively more important after the breakout emission fades.
The wind shock, by contrast, remains non-relativistic, with $\beta_{\rm sh}\simeq0.055$--$0.065$, and shows a mild deceleration followed by re-acceleration. This behaviour results from the competition between ejecta mass loading and acceleration through pressure work powered by the injected magnetar spin-down energy. At early times, the inertia of the swept-up ejecta and the conversion of bulk kinetic energy into shocked internal energy temporarily dominate, causing $\beta_{\rm sh}$ to decrease. Once the shock enters the steeper outer ejecta density profile, the density and mass-loading rate decline rapidly, allowing the pressure work associated with the injected energy to drive the shock to re-accelerate. The jet component does not show a comparable deceleration because its effective ejecta mass is much smaller and the spin-down power is concentrated towards the polar region, resulting in a larger energy input per unit swept-up mass. 
The wind bolometric emission evolves on a longer timescale and reaches a later, broader maximum at $t_{\rm obs}\sim10^{4}$--$10^{5}\,{\rm s}$. Its magnetar-powered SN component also rises more gradually and contributes significantly to the late-time bolometric luminosity. The different dynamical and diffusion timescales of the two components thus lead to distinct observational roles: the jet component mainly accounts for the early thermal X-ray emission, whereas the wind component produces the broader and later emission required to reproduce the UV/optical light curves.

The optical depths shown in Fig.~\ref{dynamic evolution} include the opacity of the ionized ejecta but not an additional contribution from electron--positron pairs. Pair loading is expected to be negligible for the non-relativistic wind shock. For the jet component, the relevant shock velocity is that relative to the expanding upstream ejecta,
$\beta_{\rm rel}=(\beta_{\rm sh}-\beta_{\rm ej})/(1-\beta_{\rm sh}\beta_{\rm ej}).$
Although the source-frame shock velocity remains at $\beta_{\rm sh}\simeq0.62$--$0.67$, the corresponding relative velocity decreases from $\beta_{\rm rel}\simeq0.31$ at early times to $\simeq0.08$ at $t_{\rm obs}\sim10^{4}\,{\rm s}$. Substantial pair loading is therefore unlikely to dominate the opacity in the present model. However, the pair abundance also depends on the local thermodynamic and radiative conditions within the radiation-mediated shock transition, which are not resolved in the present semi-analytic treatment.

\subsection{Multiband fit and inferred physical properties of GRB~060218}
\label{sub3:2}
\subsubsection{Multiband reproduction of the thermal emission}
\begin{figure}[!t]
	\centering
	\includegraphics[width=\columnwidth]{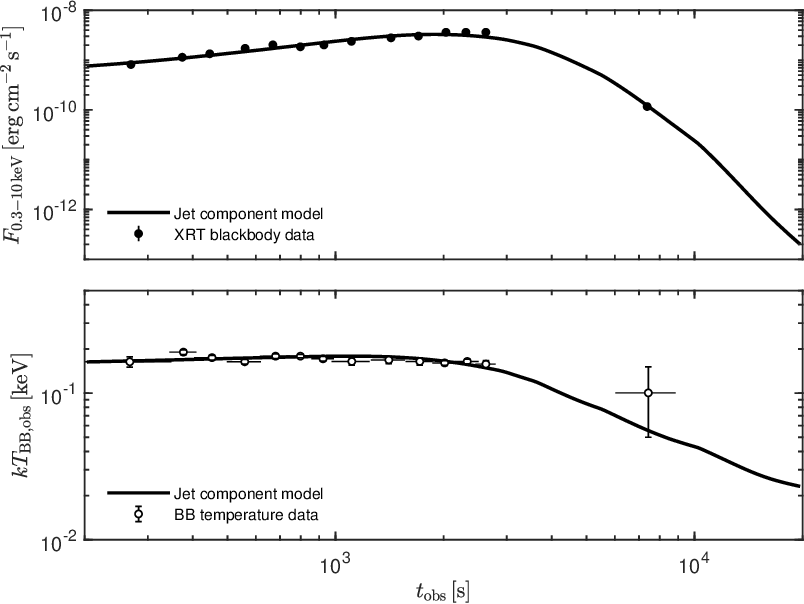}
	
	\caption{
		Best fit to the X-ray blackbody component of GRB~060218 using the jet
		component of our model. Top: $0.3$--$10\,{\rm keV}$ flux of the X-ray blackbody component.
		Bottom: Equivalent observer-frame blackbody temperature, obtained by fitting the summed spectra from all angular annuli with a
		single-temperature blackbody in the X-ray band.
		The best-fit parameter values are listed in
		Table~\ref{tab:grb060218_jet_fit}. The observational data were taken from
		\citet{2006Natur.442.1008C}.
	}
	
	\label{jet fit}
\end{figure}

\begin{figure*}[!t]
	\centering
	\includegraphics[width=0.95\textwidth]{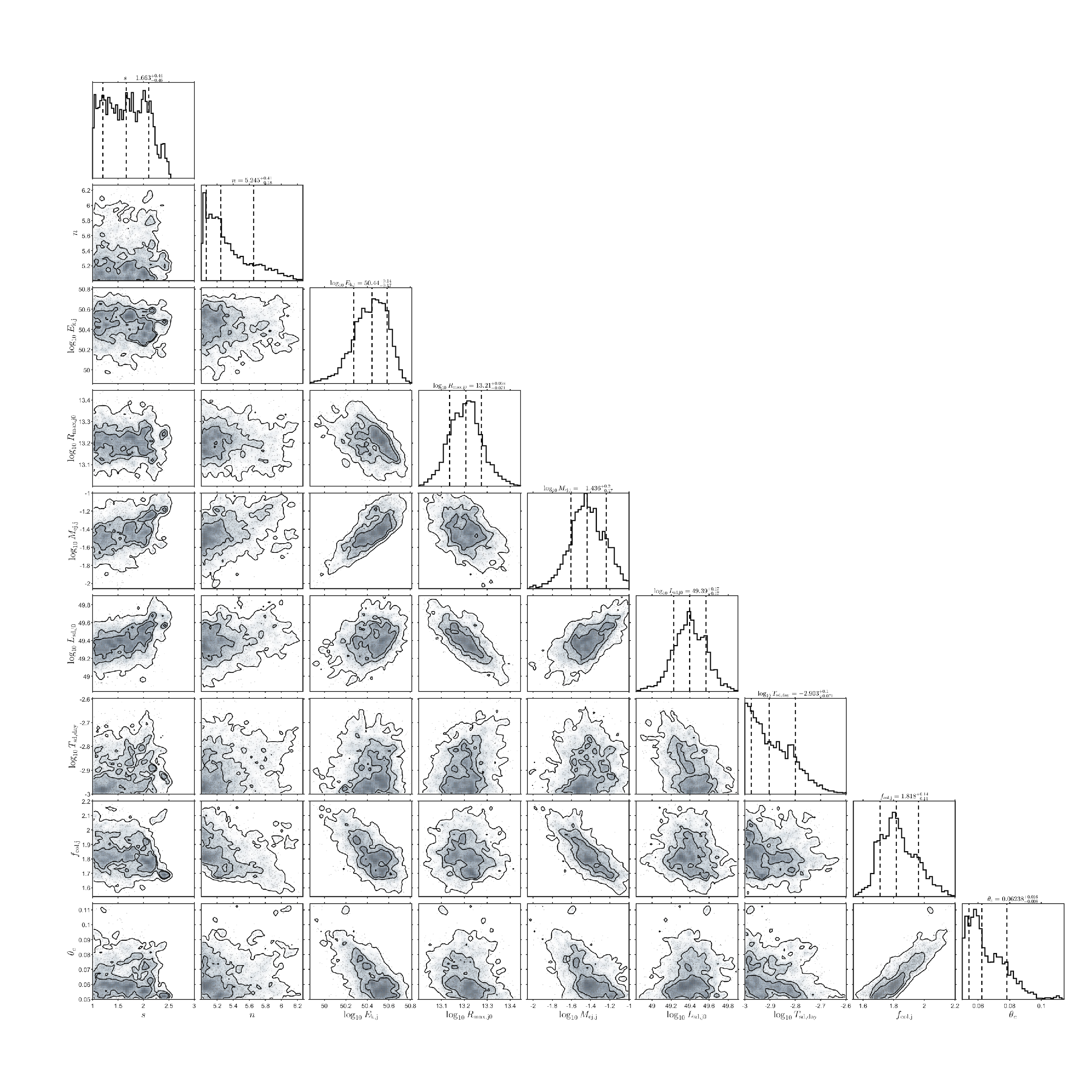}
	
	\caption{
		Posterior distributions of the model parameters, obtained from fitting the X-ray blackbody light curve of GRB~060218, and the corresponding evolution of the blackbody temperature, with the jet-component model described in Sect.~\ref{sub2:1}.
	}
	
	\label{jet corner}
\end{figure*}

The resulting best-fit jet model is shown in Fig.~\ref{jet fit}. The top panel shows that the jet component reproduces the observed $0.3$--$10\,{\rm keV}$ flux of the X-ray blackbody component from a few hundred seconds to several $10^{3}\,{\rm s}$. The model flux rises gradually, reaches a broad maximum at $t_{\rm obs}\sim10^{3}$--$3\times10^{3}\,{\rm s}$, and then declines rapidly after several $10^{3}\,{\rm s}$, broadly following the observed evolution of the thermal X-ray component. The bottom panel shows that the same model also reproduces the nearly constant early-time blackbody temperature, $kT_{\rm obs}\sim0.17\,{\rm keV}$, followed by its subsequent decline. The simultaneous agreement with the X-ray blackbody flux and equivalent blackbody temperature indicates that the polar jet component can reproduce the main observed properties of the early thermal X-ray emission. The posterior distributions of the jet parameters are shown in Fig.~\ref{jet corner}, and the corresponding constraints are summarized in Table~\ref{tab:grb060218_jet_fit}. Given the relatively large number of free parameters compared with the X-ray data points, some parameters may not be uniquely constrained by the current data.

\begin{figure}[!t]
	\centering
	\includegraphics[width=\columnwidth]{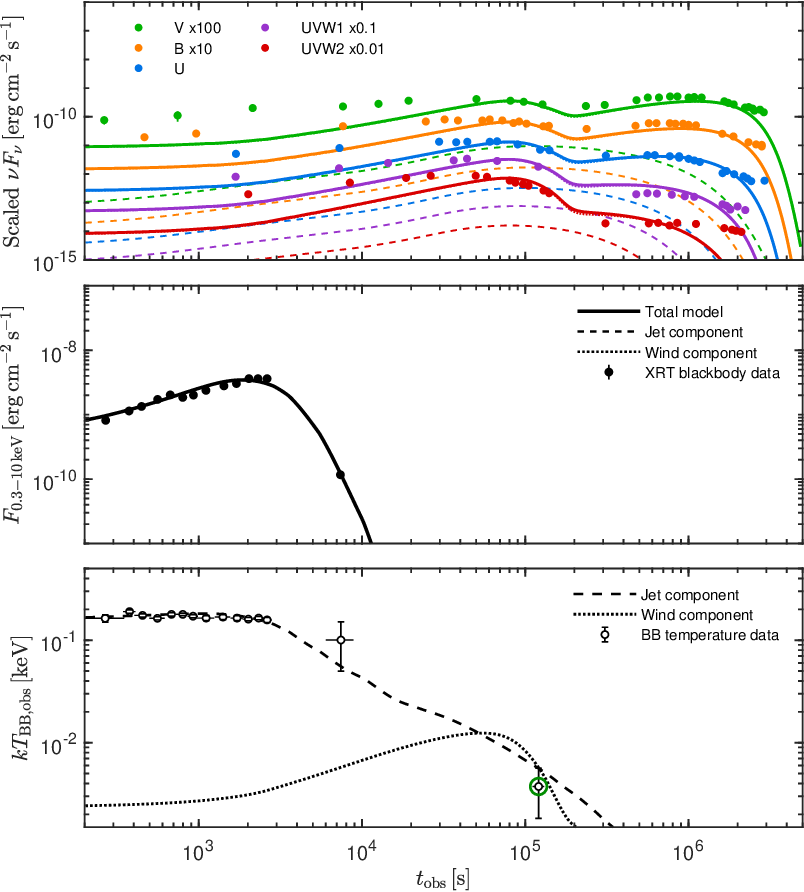}
	
	\caption{
		Best fit to the multiwavelength emission of GRB~060218, obtained with the
		combined jet--wind model. The jet contribution was fixed to the best-fit model
		inferred from the X-ray blackbody flux and temperature evolution, while the
		wind-component parameters were fitted and are listed in Table~\ref{tab:grb060218_wind_fit}. The dashed and dotted lines denote the jet and wind contributions, respectively, and the solid lines denote their sum.
		Top: Scaled UV/optical $\nu F_{\nu}$ light curves together
		with the corresponding model fits. Different colours represent different UVOT
		bands, and the individual light curves are multiplied by the factors
		indicated in the legend for clarity. Middle:
		$0.3$--$10\,{\rm keV}$ flux of the X-ray blackbody component and the corresponding model
		fit. Bottom: Corresponding blackbody-temperature evolution.
		The temperature data before \(10^4\,{\rm s}\) were derived from the XRT blackbody
		component, whereas the last point, circled in green, was derived from the UVOT
		component. The observational data were taken from
		\citet{2006Natur.442.1008C} and \citet{2015ApJ...807..172N}.
	}
	\label{full fit}
\end{figure}

The combined jet--wind fit is shown in Fig.~\ref{full fit}. The top panel presents the scaled UV/optical light curves together with the contributions from the two components. Over the epochs successfully reproduced by the model, the UV/optical emission is dominated by the wind component, while the fixed jet component contributes negligibly in these bands. The wind component reproduces the broad optical evolution from $t_{\rm obs}\sim10^{4}\,{\rm s}$ through the double-peaked phase observed after $t_{\rm obs}\gtrsim3\times10^{4}\,{\rm s}$.
The earliest UV/optical data at $t_{\rm obs}\lesssim10^{4}\,{\rm s}$ are not well reproduced by the present thermal jet--wind model, indicating that the early UV/optical excess cannot be fully accounted for by the thermal components considered here. Previous studies have likewise shown that the early UV/optical and X-ray emission of GRB~060218 cannot be consistently described by a single blackbody component (\citealp{2007MNRAS.375L..36G}). In addition, \citet{2019MNRAS.484.5484E} suggest that the optical emission around $\sim10^{3}\,{\rm s}$ may contain a synchrotron or other nonthermal contribution. Additional spectral components or departures from a Planck spectrum in the SBO emission may therefore be required at these early epochs (\citealp{2025MNRAS.542.1269I}).
At the latest epochs, the thermal jet--wind model declines more rapidly than the observed UV/optical emission, which exhibits a shallower, approximately power-law tail. As the ejecta expand and their optical depth decreases, the reservoir of trapped thermal energy becomes progressively depleted and the ejecta become less effective at storing and reprocessing the deposited energy. The resulting late-time decline in the present grey thermal treatment is therefore steeper than observed. The more slowly declining observed tail may indicate an additional emission component that is not included in the present model. In particular, as the ejecta become optically thin, nonthermal emission from the pulsar wind nebula may escape without being fully thermalized and contribute to the late-time UV/optical emission (\citealp{2022ApJ...936...54Z}).

\begin{figure*}[!t]
	\centering
	\includegraphics[width=0.8\textwidth]{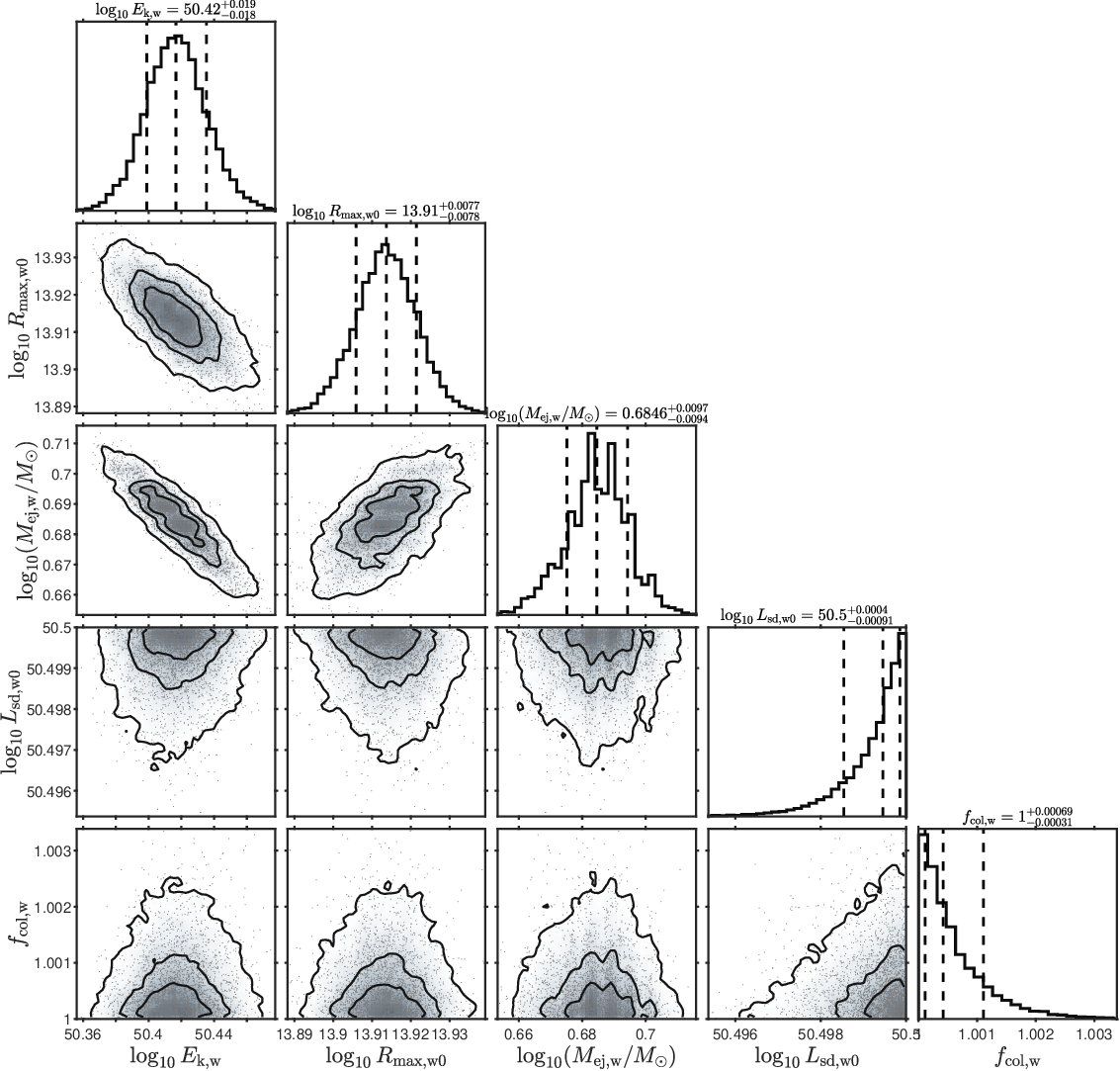}
	
	\caption{
		Posterior distributions of the wind-component parameters inferred from
		fitting the multiwavelength emission of GRB~060218 with the combined
		jet--wind model. In this fit, the jet contribution was fixed to the best-fit
		model obtained from the X-ray blackbody flux and blackbody-temperature
		evolution, while varying only the wind-component parameters.
	}
	\label{full corner}
\end{figure*}

The middle and bottom panels of Fig.~\ref{full fit} show the corresponding X-ray blackbody flux and temperature evolution. The X-ray blackbody emission remains dominated by the jet component, with a negligible contribution from the wind in the $0.3$--$10\,{\rm keV}$ band. The jet component reproduces the early XRT blackbody temperatures of $kT_{\rm obs}\sim0.17\,{\rm keV}$ before $10^{4}\,{\rm s}$, whereas the wind component has a much lower characteristic temperature, reaching only $kT_{\rm obs,w}\sim10^{-2}\,{\rm keV}$. This lower temperature is consistent with its dominant contribution in the UV/optical bands. The combined model therefore associates the early soft X-ray thermal emission primarily with the polar jet component and the later UV/optical emission with the quasi-isotropic magnetar wind. The posterior distributions of the wind-component parameters are shown in Fig.~\ref{full corner}, and the corresponding constraints are summarized in Table~\ref{tab:grb060218_wind_fit}. The posterior distributions of $L_{\rm sd,w0}$ and $f_{\rm col,w}$ reach the adopted prior boundaries, and their quoted values should therefore not be interpreted as tightly determined two-sided constraints.

\subsubsection{Inferred ejecta and engine properties}
We further compared the ejecta properties inferred for the jet and wind components. For the jet component, the relevant solid angle is estimated as $\Omega_{\rm j}=2\pi(1-\cos\theta_{\rm max})$, whereas the wind component is normalized to $4\pi$. Using the best-fit parameters listed in Tables~\ref{tab:grb060218_jet_fit} and \ref{tab:grb060218_wind_fit}, the total ejecta mass associated with the wind component is substantially larger than that associated with the jet component. After accounting for their different solid angles, however, the ejecta masses per unit solid angle are comparable. The kinetic-energy distribution shows a stronger angular contrast. Although the total kinetic energies of the two components are of the same order of magnitude, the much smaller solid angle of the jet component implies a larger kinetic energy per unit solid angle and a higher specific kinetic energy in the polar ejecta. This corresponds to a higher characteristic velocity scale in the polar region, consistent with the faster jet-driven shock shown in Fig.~\ref{dynamic evolution}. These quantities should be interpreted as effective properties of the jet-interacting and wind-interacting ejecta within the present two-component model, rather than as a direct reconstruction of the global angular structure of the SN ejecta.
The fitted value $n=5.149$ indicates a relatively shallow outer ejecta density profile, implying that a non-negligible fraction of the kinetic energy is contained in the fast-moving ejecta. Similar high-velocity ejecta structures have been obtained in simulations of engine-driven and choked-jet explosions (\citealp{2022MNRAS.517..582E,2022ApJ...925..148S,2023MNRAS.519.1941P,2024PASJ...76..863S}) and are also favoured by SBO interpretations of GRB~060218 (\citealp{2015ApJ...807..172N,2025MNRAS.542.1269I}). The relatively shallow density profile inferred here may therefore reflect a more general property of engine-driven explosions rather than a feature specific to the present model.

The best-fit engine parameters further indicate that the spin-down luminosity assigned to the jet direction is smaller than that supplied to the wind component. Within the present two-component parametrization, this corresponds to a larger effective spin-down power input into the quasi-isotropic wind than into the polar jet.

\subsubsection{Sensitivity to radiative and spectral assumptions}
For the jet component, the fitted colour-correction factor $f_{\rm col,j}>1$ indicates that the characteristic colour temperature inferred from the X-ray spectrum is higher than the effective temperature predicted by the grey-opacity diffusion model. Such a difference is expected when the emergent radiation is not fully thermalized at the
effective photosphere. For example, in a scattering-dominated medium, photons may thermalize at deeper and hotter layers before escaping.
Additional frequency-dependent transfer effects, including Comptonization, may further modify the relation between the effective and colour temperatures. The wind component is constrained mainly by the UV/optical emission, for which the inferred spectrum is less sensitive to this temperature correction. 

The colour-corrected blackbody adopted in the baseline model should be regarded as an effective spectral prescription rather than a self-consistent calculation of the spectrum emerging from a radiation-mediated shock.
In sufficiently fast shocks, inefficient photon production and incomplete thermalization may drive the radiation away from a Planck spectrum, while Comptonization can further modify the characteristic radiation temperature and spectral shape
(\citealp{2010ApJ...725..904N,2010ApJ...716..781K,2013ApJ...774...79S,2022ApJ...928..122M,2025MNRAS.542.1269I}).
Monte Carlo calculations of fast and relativistic radiation-mediated shocks likewise predict spectra that are substantially flatter than a Planck spectrum below the spectral peak
(\citealp{2020MNRAS.499.4961I,2026MNRAS.547ag389I}).
The thermalization conditions may also differ among angular annuli.
We therefore performed two post-processing sensitivity tests using the same best-fit dynamical solution.

\begin{figure}[!t]
	\centering
	\includegraphics[width=\columnwidth]{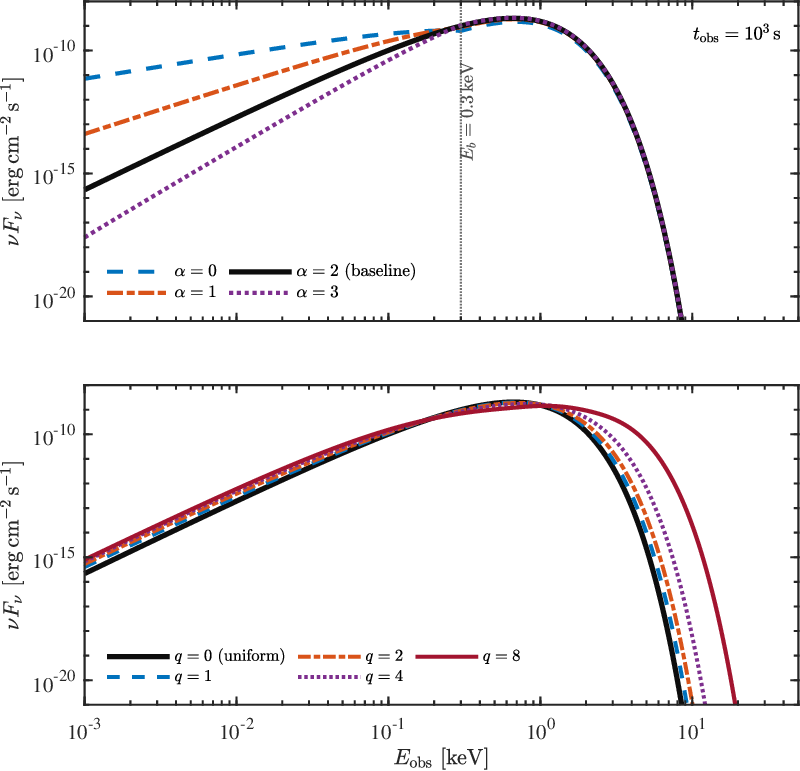}
	
	\caption{
		Sensitivity of the jet spectrum at $t_{\rm obs}=10^{3}\,{\rm s}$, calculated using the same best-fit dynamical solution.
		Top: Spectrum below $E_{\rm b}=0.3\,{\rm keV}$, modified by
		$(\nu/\nu_{\rm b})^{\alpha-2}$ for $\alpha=0$, 1, 2, and 3 and
		renormalized in each annulus to preserve its bolometric luminosity.
		Here, $\alpha=2$ recovers the baseline Rayleigh--Jeans behaviour.
		Bottom: Spectra obtained with $f_{{\rm col},i}=1+(f_{\rm col,j}-1) (\beta_{{\rm rel},i}/\beta_{\rm ref})^{q}$, where $\beta_{\rm ref}$ is the luminosity-weighted mean relative shock velocity of the emitting annuli. Here, $q=0$ recovers the uniform-$f_{\rm col}$ baseline.
		Flatter low-energy spectra enhance the UV/optical emission, while stronger velocity dependence of $f_{\rm col,j}$ broadens the integrated spectrum.
	}
	
	\label{fig:spectral_sensitivity}
\end{figure}

First, we modified the spectrum of each annulus according to
\begin{equation}
	F_{\nu,i}^{(\alpha)}
	=
	C_{\alpha,i}F_{\nu,i}^{(0)}
	\begin{cases}
		(\nu/\nu_{\rm b})^{\alpha-2}, & \nu<\nu_{\rm b},\\
		1, & \nu\geq\nu_{\rm b},
	\end{cases}
\end{equation}
where $F_{\nu,i}^{(0)}$ is the baseline spectrum, $E_{\rm b}=h\nu_{\rm b}=0.3\,{\rm keV}$, corresponding to the lower boundary of the XRT band, and $C_{\alpha,i}$ was chosen to preserve the bolometric luminosity of each annulus. We considered $\alpha=0$, 1, 2, and 3, with $\alpha=2$ exactly recovering the baseline Rayleigh--Jeans behaviour. 
These values were intended to bracket possible departures from a Planck spectrum rather than to represent self-consistent radiative-transfer solutions.
As shown in the top panel of Fig.~\ref{fig:spectral_sensitivity}, flatter low-energy spectra ($\alpha<2$) substantially enhance the UV/optical emission while producing only minor changes around the X-ray spectral peak.
A non-Planckian low-energy spectrum could therefore partly alleviate the early optical deficit of the baseline model, although a quantitative prediction requires frequency-dependent radiative transfer.

Second, we explored the possible effect of annulus-dependent thermalization by adopting
\begin{equation}
	f_{{\rm col},i}
	=
	1+
	(f_{\rm col,j}-1)
	\left(
	\frac{\beta_{{\rm rel},i}}
	{\beta_{\rm ref}}
	\right)^q ,
\end{equation}
where $\beta_{{\rm rel},i}$ is the shock velocity relative to the upstream ejecta and
\begin{equation}
	\beta_{\rm ref}
	=
	\frac{\sum_i L_{{\rm rad},i}\beta_{{\rm rel},i}}
	{\sum_i L_{{\rm rad},i}}
\end{equation}
is the luminosity-weighted mean relative shock velocity of the emitting annuli. 
For the best-fit solution at $t_{\rm obs}=10^{3}\,{\rm s}$, $\beta_{\rm ref}\simeq0.18$.
The case $q=0$ exactly recovers the uniform-$f_{\rm col,j}$ baseline, while $q=1$, 2, 4, and 8 progressively increase the sensitivity of $f_{\rm col,j}$ to the relative shock velocity. These values were used only as phenomenological sensitivity tests and are not intended as a physical scaling of $f_{\rm col,j}$ with shock velocity. 
In particular, the $q=8$ case is deliberately extreme and should not be identified with the steep dependence of the emergent radiation temperature on shock velocity found in non-equilibrium SBO calculations.

As shown in the bottom panel of Fig.~\ref{fig:spectral_sensitivity}, increasing the velocity dependence of $f_{\rm col,j}$ progressively broadens the angle-integrated spectrum. 
This demonstrates that variations in the thermalization conditions among angular annuli can broaden the multi-temperature jet emission. Even in the deliberately extreme $q=8$ case, however, the spectrum remains strongly curved rather than forming an extended power law over a broad energy range. 
We therefore did not identify this effect with the observed nonthermal X-ray component of GRB~060218. We retained a single $f_{\rm col,j}$ in the baseline fit and interpret it as an effective colour correction for the integrated thermal component, rather than implying identical local thermalization conditions in all annuli. 
A self-consistent determination of the spectral shape and its angular dependence requires frequency-dependent radiation-transfer calculations and is beyond the scope of the present model.

\section{Conclusions and discussion}
\label{sec4}

In this work, we developed a two-component magnetar-powered model to interpret the thermal radiative components of GRB~060218--SN~2006aj. Motivated by the possible coexistence of a collimated polar jet and a quasi-isotropic magnetar wind (\citealp{2015MNRAS.454.3311M,2018MNRAS.475.2659M, 2018ApJ...862..130L,2019ApJ...877..153Z}), we phenomenologically divided the spin-down power of the central NS into a jet component and a wind component. Both components interact with the surrounding SN ejecta and drive forward shocks, producing SBO emission and subsequent magnetar-powered thermal emission. We applied this model to the X-ray blackbody component and the UV/optical light curves of GRB~060218--SN~2006aj.

We first fitted the $0.3$--$10\,{\rm keV}$ flux of the X-ray blackbody component and the corresponding blackbody-temperature evolution using only the jet component. In this calculation, the jet was divided into angular annuli, and the model temperature compared with the observations was obtained by fitting a single-temperature blackbody to the summed observer-frame spectrum from all annuli. The resulting equivalent blackbody temperature, together with the X-ray blackbody flux, can be reproduced by the jet component from a few hundred seconds to several $10^{3}\,{\rm s}$. The corresponding dynamical evolution shows that the jet-driven shock in the X-ray-emitting polar region reaches mildly relativistic velocities, with $\beta_{\rm sh}\simeq0.62$--$0.67$. The bolometric luminosity of the jet component is initially dominated by the SBO contribution, while the magnetar-powered SN component declines more slowly and becomes relatively more important after the breakout emission fades.

We then fixed the best-fit jet component and added the wind component to fit the UV/optical light curves. In contrast to the X-ray blackbody emission, which is dominated by the jet component, the UV/optical emission is mainly reproduced by the wind component. The wind component reproduces the broad optical evolution from $t_{\rm obs}\sim10^{4}\,{\rm s}$ to late times and dominates the late-time optical light curves. The inferred properties of the wind component are conditional on the adopted host-galaxy extinction, since a lower host extinction would yield a dereddened UV/optical spectrum that is less consistent with a purely thermal spectrum. The wind shock remains non-relativistic, with $\beta_{\rm sh}\simeq0.055$--$0.065$, and its bolometric emission evolves on a longer timescale than that of the jet component. The different dynamical timescales of the two components therefore provide a consistent description of the early soft X-ray thermal emission and the broader UV/optical evolution.

The early X-ray emission of GRB~060218 is dominated by a broad nonthermal component, while the blackbody component considered here contributes only a subdominant fraction of the total X-ray output (\citealp{2006Natur.442.1008C,2007ApJ...654..385K}). Its physical origin remains uncertain, with proposed interpretations including synchrotron or inverse-Compton emission associated with a mildly relativistic outflow, bulk Comptonization of breakout photons, and non-equilibrium SBO emission (\citealp{2006astro.ph..4510D,2007MNRAS.375L..36G,2007ApJ...664.1026W,2016MNRAS.460.1680I,2025MNRAS.542.1269I}). Within our framework, particle acceleration at the jet-driven forward shock could in principle contribute to this nonthermal emission through synchrotron and inverse-Compton processes. These effects were not included here: our analysis was restricted to the blackbody flux and temperature inferred from the power-law-plus-blackbody spectral decomposition. A quantitative treatment of the nonthermal component would require additional particle-acceleration and frequency-dependent radiative-transfer calculations.

The earliest UV/optical data are not fully reproduced within the present thermal jet--wind framework, suggesting that additional spectral effects or emission components may contribute at these epochs. As shown by the spectral sensitivity tests above, departures from a Planck spectrum associated with incomplete thermalization and spectral modification in the shock-heated ejecta can enhance the low-energy emission relative to the baseline blackbody prescription and may therefore partially account for the early UV/optical excess.
Previous studies have also shown that a single blackbody component cannot simultaneously account for the early UV/optical and X-ray emission of GRB~060218 (\citealp{2007MNRAS.375L..36G}). The early UV/optical excess may also reflect additional emission, such as synchrotron radiation from external shocks (\citealp{2019MNRAS.484.5484E}) or free-free emission in a non-equilibrium SBO scenario (\citealp{2025MNRAS.542.1269I}). At later times, as the ejecta become optically thin, leakage of nonthermal radiation from the pulsar wind nebula may provide an additional contribution to the residual UV/optical emission (\citealp{2022ApJ...936...54Z}). These additional emission components are not included self-consistently in the present thermal model.

The best-fit parameters suggest different effective ejecta properties for the jet- and wind-interacting components. The wind component is associated with a much larger total ejecta mass, whereas the mass loading per unit solid angle becomes comparable to that of the jet component after accounting for the
different solid angles. The polar ejecta, however, have a larger kinetic energy per unit solid angle and per unit ejecta mass. 
The best-fit parameters give $L_{\rm sd,w0}>L_{\rm sd,j0}$, suggesting that the effective initial spin-down power carried by the quasi-isotropic wind exceeds that carried by the polar jet.
The fitted colour-correction factors also differ between the two components. For the jet component, $f_{\rm col,j}>1$ implies that the colour temperature inferred from the X-ray spectrum exceeds the effective photospheric temperature predicted by the model. By contrast, $f_{\rm col,w}\simeq1$ indicates that the UV/optical data do not require a substantial colour correction.

Several simplifying assumptions should be noted. First, in a realistic multidimensional system, the polar jet and the quasi-isotropic magnetar wind are unlikely to remain completely independent. Pressure coupling through the magnetar-wind nebula may redistribute energy and momentum across angle (\citealp{2008MNRAS.383L..25B,2009MNRAS.396.2038B}), while shear and interface instabilities may produce mixing near the jet boundary (\citealp{2021MNRAS.500.3511G,2023MNRAS.520.3009A}). In the present semi-analytic model, the two components are evolved separately to isolate their leading radial dynamical and radiative contributions. This decomposition should therefore be regarded as an effective modelling approximation rather than as complete physical decoupling. Second, in analogy with previous semi-analytic treatments of structured jets (\citealp{2017MNRAS.472.4953L,2019A&A...628A..18S}), the jet region is divided into angular annuli that evolve independently in a quasi-1D approximation. Lateral transport of momentum and internal energy between neighbouring annuli is neglected. The annulus-dependent quantities should therefore be interpreted as effective angular properties within this approximation.

Third, the shock velocity and the bulk motion of the shocked ejecta in each annulus are assumed to be radial. In a nonspherical jet--ejecta interaction, the local shock normal may deviate from the radial direction, particularly near the jet boundary. Because the modelled X-ray emission is concentrated at small polar angles, radial propagation is adopted as a leading-order approximation. Non-radial motion may nevertheless modify the detailed angular evolution and the inferred annulus-dependent parameters. Fourth, the optical depth of the unshocked ejecta, $\tau_{\rm un}$, is evaluated along the radial direction, and the diffusion timescale is estimated assuming predominantly radial photon transport. Non-radial escape paths are not included and could modify the effective diffusion time and emergent luminosity.

Fifth, the present calculations assume an on-axis observer. For an off-axis viewing angle, the Doppler boosting of the polar ejecta would be reduced and the relative contributions and arrival times of different angular annuli would change. The thermal emission would therefore tend to appear fainter and softer, with a broader or later light-curve peak. A quantitative treatment of arbitrary viewing angles requires integration over the corresponding equal-arrival-time surfaces and is beyond the scope of the present work. A more complete treatment of the simplifying assumptions and additional physical effects discussed above will require multidimensional relativistic magnetohydrodynamic simulations coupled with detailed radiation-transfer calculations.

Although developed for GRB~060218, the present model may also be relevant to other engine-driven type Ic-BL explosions in which central-engine activity gives rise to detectable thermal emission through its interaction with the ejecta. GRB~100316D--SN~2010bh is one of the closest Swift analogues of GRB~060218, sharing its long duration, low luminosity, soft spectrum, and a reported early thermal X-ray component (\citealp{2011ApJ...726...32F,2011MNRAS.411.2792S,2012MNRAS.427.2950S}). GRB~171205A--SN~2017iuk also shows evidence of thermal X-ray emission, although the physical origin of this component remains uncertain (\citealp{2018A&A...619A..66D,2021MNRAS.501.4974V}). More recently, the associations of EP~240414a--SN~2024gsa and EP~250108a--SN~2025kg with type Ic-BL SNe have broadened the sample of soft X-ray transients associated with engine-driven stellar explosions (\citealp{2025NatAs...9.1073S,2025ApJ...988L..13R,2025ApJ...988L..60S}). Their multiwavelength emission has been discussed in connection with weak relativistic or choked jets, cocoon emission, and interaction with circumstellar material (\citealp{2025ApJ...986L...4H,2025ApJ...985...21Z,2025ApJ...988L..60S,2025ApJ...988L..13R}). Although these events do not necessarily provide direct tests of the thermal jet--wind model, they offer useful comparison cases for assessing the broader role of central-engine activity in soft X-ray transients associated with type Ic-BL SNe. The present model is not intended to reproduce the full diversity of low-luminosity GRBs and fast X-ray transients.
Nevertheless, it is noteworthy that a significant fraction of well-observed low-luminosity GRBs and related X-ray transients show evidence for an excess soft X-ray component with a characteristic temperature of order $kT\sim0.1$~keV, even when this thermal component is subdominant to the nonthermal emission (\citealp{2025MNRAS.542.1269I}).
Events whose observed properties are dominated by nonthermal jet or cocoon emission, strong circumstellar interaction, or viewing-angle effects may require physical ingredients beyond those included here. In particular, events for which a thermal component cannot be robustly isolated spectroscopically do not yet provide a direct test of the present model.

In summary, the combined jet--wind model provides a physically motivated framework for interpreting the thermal emission of GRB~060218--SN~2006aj. In this picture, the early soft X-ray blackbody component is associated with the polar jet--ejecta interaction, while the broader UV/optical light curves are mainly powered by the magnetar wind interacting with a more massive ejecta component. Although additional nonthermal or non-equilibrium processes are likely required to explain the earliest UV/optical excess, the model supports a two-component energy-injection scenario in which a newborn magnetar deposits energy into both a collimated jet and a quasi-isotropic wind. Such a configuration provides a plausible framework for understanding how magnetar activity may shape the multiwavelength emission of some low-luminosity GRB--SN systems.

\begin{acknowledgements}
	We thank the anonymous referee for helpful feedback on the manuscript. This work is supported by the National Natural Science Foundation of China (grant Nos. 12494575, 12273005, and 12133003), the special funding for Guangxi Bagui Youth Scholars (Da-Bin Lin), the National Key R\&D Program of China (grant No. 2024YFA1611700), the Guangxi Talent Program (“Highland of Innovation Talents”), and the Innovation Project of Guangxi Graduate Education (grant No. YCBZ2026032).

\end{acknowledgements}
\bibliography{sample631}

\end{document}